\documentclass[fleqn,usenatbib]{mnras}

\usepackage[modulo]{lineno}
\usepackage{ulem}
\usepackage{soul,xcolor}
\usepackage{xcolor}
\newcommand{\teff}{$T_{\rm eff}$} 
\newcommand{\logg}{$\log g$} 
\newcommand{\kms}{km s$^{-1}$}
\newcommand{\vt}{$\xi_t$} 
\newcommand{\fei}{Fe\,{\sc i}}
\newcommand{\feii}{Fe\,{\sc ii}}
\newcommand{\nai}{Na\,{\sc i}}
\newcommand{\mgi}{Mg\,{\sc i}}
\newcommand{\ali}{Al\,{\sc i}}
\newcommand{\sii}{Si\,{\sc i}}
\newcommand{\cai}{Ca\,{\sc i}}
\newcommand{\tii}{Ti\,{\sc i}}
\newcommand{\tiii}{Ti\,{\sc ii}}
\newcommand{\scii}{Sc\,{\sc ii}}
\newcommand{\cri}{Cr\,{\sc i}}
\newcommand{\mni}{Mn\,{\sc i}}
\newcommand{\coi}{Co\,{\sc i}}
\newcommand{\nii}{Ni\,{\sc i}}
\newcommand{\zni}{Zn\,{\sc i}}
\newcommand{\srii}{Sr\,{\sc ii}}
\newcommand{\yii}{Y\,{\sc ii}}
\newcommand{\zrii}{Zr\,{\sc ii}}
\newcommand{\baii}{Ba\,{\sc ii}}
\newcommand{\euii}{Eu\,{\sc ii}}
\newcommand{\lii}{Li\,{\sc i}}

\usepackage[T1]{fontenc}

\DeclareRobustCommand{\VAN}[3]{#2}
\let\VANthebibliography\thebibliography
\def\thebibliography{\DeclareRobustCommand{\VAN}[3]{##3}\VANthebibliography}

\usepackage{graphicx}	
\usepackage{amsmath}	
\usepackage{amssymb}	

\usepackage{longtable}

\title[SkyMapper DR1.1 Abundances]{{High resolution spectroscopic follow-up of the most metal-poor candidates from SkyMapper DR1.1}\thanks{This paper includes data gathered with the 6.5 m Magellan Telescopes located at Las Campanas Observatory, Chile. Some of the data presented herein were obtained at the W.\ M.\ Keck Observatory, which is operated as a scientific partnership among the California Institute of Technology, the University of California and the National Aeronautics and Space Administration. The Observatory was made possible by the generous financial support of the W.\ M.\ Keck Foundation.}}

\author[D. Yong et al.]
{D. Yong,$^{1,2}$\thanks{E-mail: david.yong@anu.edu.au}
G. S. Da Costa,$^{1,2}$
M. S. Bessell,$^{1}$
A. Chiti,$^{3}$
A. Frebel,$^{3}$
X. Gao,$^{4}$ \newauthor 
K. Lind,$^{5}$
A. D. Mackey,$^{1}$
A. F. Marino,$^{6}$
S. J. Murphy,$^{7}$
T. Nordlander,$^{1,2}$
M. Asplund, \newauthor
A. R. Casey,$^{8,2}$
C. Kobayashi,$^{9,2}$
J. E. Norris$^{1}$ and 
B. P. Schmidt$^{1}$
\\
$^{1}$Research School of Astronomy and Astrophysics, Australian National University, Canberra, ACT 0200, Australia\\
$^{2}$ARC Centre of Excellence for Astrophysics in Three Dimensions (ASTRO-3D), Australia\\
$^{3}$Department of Physics and Kavli Institute for Astrophysics and Space Research, Massachusetts Institute of Technology, Cambridge, MA 02139, USA\\
$^{4}$Max-Planck-Institut f{\" u}r Astronomie (MPIA), K{\" o}nigstuhl 17, D-69117 Heidelberg, Germany\\
$^{5}$Department of Astronomy, Stockholm University, AlbaNova University Center, 106 91 Stockholm, Sweden\\
$^{6}$Istituto Nazionale di Astrofisica - Osservatorio Astronomico di Arcetri, Largo Enrico Fermi, 5, 50125, Firenze, Italy\\
$^{7}$School of Science, The University of New South Wales, Canberra, ACT 2600, Australia\\
$^{8}$School of Physics \& Astronomy, Monash University, Wellington Road, Clayton 3800, Victoria, Australia\\
$^{9}$Centre for Astrophysics Research, Department of Physics, Astronomy and Mathematics, University of Hertfordshire, Hatfield, AL10 9AB, UK\\
}

\date{Accepted XXX. Received YYY; in original form ZZZ}

\pubyear{2021}

\begin{document}
\setstcolor{red}
\label{firstpage}
\pagerange{\pageref{firstpage}--\pageref{lastpage}}
\maketitle

\begin{abstract}
We present chemical abundances for 21 elements (from Li to Eu) in 150 metal-poor Galactic stars spanning $-$4.1 $<$ [Fe/H] $<$ $-$2.1. The targets were selected from the SkyMapper survey and include 90 objects with [Fe/H] $\le$ $-$3 of which some 15 have [Fe/H] $\le$ $-$3.5. When combining the sample with our previous studies, we find that the  metallicity distribution function has a power-law slope of $\Delta$(log N)/$\Delta$[Fe/H] = 1.51 $\pm$ 0.01 dex per dex over the range $-$4 $\le$ [Fe/H] $\le$ $-$3. With only seven carbon-enhanced metal-poor stars in the sample, we again find that the selection of metal-poor stars based on SkyMapper filters is biased against highly carbon rich stars for [Fe/H] $>$ $-$3.5. Of the 20 objects for which we could measure nitrogen, 11 are nitrogen-enhanced metal-poor stars. Within our sample, the high NEMP fraction (55\% $\pm$ 21\%) is compatible with the upper range of predicted values (between 12\% and 35\%). The chemical abundance ratios [X/Fe] versus [Fe/H] exhibit similar trends to previous studies of metal-poor stars and Galactic chemical evolution models. We report the discovery of nine new r-I stars, four new r-II stars, one of which is the most metal-poor known, nine low-$\alpha$ stars with [$\alpha$/Fe] $\le$ 0.15 as well as one unusual star with [Zn/Fe] = +1.4 and [Sr/Fe] = +1.2 but with normal [Ba/Fe]. Finally, we combine our sample with literature data to provide the most extensive view of the early chemical enrichment of the Milky Way Galaxy. 
\end{abstract}




\color{black}
\section{Introduction}

The most metal-poor stars in the Galaxy provide a unique opportunity to understand the Milky Way's earliest stages of formation and evolution, and the origin of the chemical elements \citep{Beers:2005aa,Frebel:2015aa}. The basic assumptions are that the metallicity of a star serves as a proxy for its age (with iron as the canonical measure of metallicity) and that the atmospheres of low mass stars retain the chemical composition of the interstellar medium at the time and place of their birth. In this context, chemical abundance studies of the most iron-poor stars probe the earliest chemical enrichment events and the properties of the previous generation(s) of stars. 

The identification and analysis of the most iron-poor stars has been a major endeavour since the discovery that some stars are metal deficient with respect to the sun \citep{Chamberlain:1951aa,Baschek:1959aa,Wallerstein:1963aa}. Important advances have come from pushing to ever lower metallicity (e.g., \citealt{Bessell:1984aa,Christlieb:2002aa,Frebel:2005aa,Caffau:2011aa,Keller:2014aa,Aguado:2018ab,Starkenburg:2018aa,Nordlander:2019aa}) in the pursuit of identifying a star whose chemical composition reflects the primordial Big Bang composition. Significant advances in our understanding of early chemical enrichment have come from studies which have sought to increase the numbers of known metal-poor stars and investigate their chemical abundance patterns (e.g., \citealt{McWilliam:1995aa,Ryan:1996aa,Johnson:2002aa,Cayrel:2004aa,Venn:2004aa,Venn:2020aa,Aoki:2006aa,Bonifacio:2009aa,Bonifacio:2012aa,Yong:2013ab,Roederer:2014ac,Hansen:2015aa,Placco:2015aa,Yoon:2016aa,Aguado:2019aa,Hansen:2019aa,Caffau:2020aa}). In parallel, theoretical efforts to study the properties and nucleosynthetic yields of the first generations of stars have been crucial in our interpretation of chemical abundance ratios in metal-poor stars (e.g., \citealt{Schneider:2003aa,Karlsson:2006aa,Salvadori:2007aa,Prantzos:2008aa,Kobayashi:2011aa,Nomoto:2013aa,Tominaga:2014aa,Clarkson:2020aa}). 

The current generation of surveys focusing on the discovery of metal-poor stars include Pristine \citep{Starkenburg:2017aa} and SkyMapper \citep{Keller:2007aa}, both of which are deep photometric surveys employing narrow-to-intermediate-band metallicity sensitive filters. As described in \citet{DaCosta:2019aa}, the "commissioning-era" of the SkyMapper survey led to the discovery of the most iron-poor star known, SMSS~J031300.36-670839.3 with [Fe/H] $<$ $-$6.5 (3D, NLTE) \citep{Keller:2014aa,Bessell:2015aa,Nordlander:2017aa}. Additional studies of metal-poor stars from the SkyMapper commissioning-era survey were reported by \citet{Jacobson:2015aa}, \citet{Howes:2016aa} and \citet{Marino:2019aa}. From the SkyMapper "main" survey, we have discovered SMSS~J160540.18-144323.1 with the lowest detected iron measurement, [Fe/H] = $-$6.2 (1D, LTE) \citet{Nordlander:2019aa}. Collectively, the results from the SkyMapper survey have provided important new data for understanding the early evolution of our Galaxy \citep[e.g.,][]{Cordoni:2020aa,Chiti2021:aa}. 

The aim of this paper is to present the high-resolution spectroscopic analysis for a sample of 150 stars selected from SkyMapper photometry which have been vetted using intermediate resolution spectroscopy on the ANU 2.3m telescope \citep[see][for details]{DaCosta:2019aa}. The paper is arranged as follows. In Section 2 we describe the sample selection, observations and data reduction. In Section 3 we present the analysis. Section 4 includes our results and our conclusions are given in Section 5. 

\section{Sample Selection, Observations and Data Reduction}

Targets were identified from the SkyMapper metallicity sensitive diagram, $m_i = (v - g)_0 - 1.5 (g-i)_0$ vs.\ $(g-i)_0$, then observed at intermediate resolution using the WiFeS \citep{Dopita:2010aa} integral field spectrograph at the ANU 2.3m telescope. Further details on the photometric selection and WiFeS spectroscopy can be found in \citet{DaCosta:2019aa}. As described in \citet{Bessell:2007aa} and \citet{Norris:2013ab}, a spectrophotometric flux fitting method was applied to the WiFeS spectra to obtain estimates of the effective temperature (\teff), surface gravity (\logg) and metallicity ([Fe/H]). 

The most promising candidates, i.e., the most metal poor based on the WiFeS spectra, were observed using the Magellan Inamori Kyocera Echelle (MIKE) spectrograph \citep{Bernstein:2003aa} at the Magellan Telescope in 2017 and 2018. Note that the WiFeS observations were continuously being obtained such that the best available targets for high-resolution observations were updated before each observing run. Depending on the observing conditions, spectra were obtained with either the 0\farcs7 or 1\farcs0 slits resulting in spectral resolutions of R = 35,000 in the blue and R = 28,000 in the red, or R = 28,000 in the blue and R = 22,000 in the red, for the smaller and larger slit sizes, respectively. The CCD binning was set to 2 $\times$ 2. Exposure times were adjusted to achieve signal-to-noise ratios of around S/N = 50 per pixel near 4500\AA. We examined the spectra at the telescope and in some instances we re-observed objects to increase the S/N. The average S/N is 54 per pixel near 4500\AA\ and the minimum and maximum values are 19 and 138, respectively. 

The spectra were reduced using the CarPy data reduction pipeline\footnote{\url{https://code.obs.carnegiescience.edu/mike}} described in \citet{Kelson:2003aa}. Multiple exposures were combined and individual orders were merged and normalised to create a single continuous spectrum per star. For the continuum normalisation, we applied two-dimensional modelling following the approach of \citet{Barklem:2002aa} and \citet{Ramirez:2008aa}. That is, we fit high order polynomials to the fluxes in each order, as well as variations in the blaze perpendicular to the dispersion. 

Note that two targets included in this study were not originally selected from the procedure described in \citet{DaCosta:2019aa}. SMSS~J054913.80-453904.0 (=HE~0547-4539) is from \citet{Barklem:2005aa} and SMSS~J143511.34-420326.4 (=SMSS~J1435-4203) is from \citet{Jacobson:2015aa}. Both objects were observed as bright back-up targets. At the time of the Magellan/MIKE observations, two of the candidates were not recognised as having published high-dispersion analyses: SMSS~J030428.44-340604.8 (=HE~0302-3417A) and SMSS~J232121.57-160505.4 (=HE~2318-1621) were studied by \citet{Hollek:2011aa} and \citet{Placco:2014ab}, respectively. We have retained them in our analysis and briefly discuss comparisons with published data at the end of Section 3. (We also observed SMSS~J100231.91-461027.5 which is a likely post-AGB star.) 

Three candidates were observed using the HIRES spectrograph \citep{Vogt:1994aa} at the Keck telescope on 02 Feb 2017. We used the red cross disperser and the C1 decker with a slit width of 0\farcs86 which provides a spectral resolution of R = 45,000. The CCD binning was 2 $\times$ 1 (spatial $\times$ spectral). The data were reduced using MAKEE\footnote{\url{https://sites.astro.caltech.edu/~tb/makee/}} and the wavelength coverage was from 4060\AA\ to 8350\AA. The S/N ratios per pixel near 4500\AA\ ranged from 29 to 79.  

The final set of 48 candidates were observed using the FEROS spectrograph \citep{Kaufer:1999aa} at the ESO 2.2m telescope in May 2018. The FEROS spectra were processed automatically using the ESO online real-time pipeline reduction. The spectral resolution was R = 48,000, CCD binning was 1 $\times$ 1 and wavelength coverage was from 3600\AA\ to 9100\AA. The S/N ranged from 5 to 79 per pixel near 4500\AA\ with an average value of 17. As we shall discuss later, we present stellar parameters (including [Fe/H]) for all stars but for the nine objects with S/N $<$ 10 we do not measure chemical abundance ratios ([X/Fe]). The program stars and observing details are presented in Table \ref{tab:obs}. 

\begin{table*}
	\caption{Program stars and observing details.}
	\label{tab:obs}
	\begin{tabular}{lccrcr}
	\hline
	ID &
	Telescope$^1$ & 
	$g$ &
	S/N$^2$ &
	JD &
	RV \\
	SMSS &
	&
	mag &
	&
	&
	(\kms) \\ 
	\hline
J001604.23-024105.0  &  M &  12.89 &   43 &  2458085.78715 &      49.3  \\  
J005420.96-844117.0  &  M &  14.59 &   38 &  2458084.99471 &     182.8  \\  
J011126.27-495048.4  &  M &  14.40 &   52 &  2458037.93753 &     255.0  \\  
J020050.19-465735.2  &  M &  13.40 &   77 &  2458035.87291 &    $-$7.0  \\  
J024246.96-470353.6  &  M &  14.90 &   42 &  2458086.04841 &       6.9  \\  
J030245.60-281454.0  &  M &  14.19 &   52 &  2458187.75078 &      45.9  \\  
J030258.53-284326.9  &  M &  14.29 &   54 &  2458188.76044 &      44.7  \\  
J030428.44-340604.8  &  M &  11.13 &  138 &  2458084.73097 &     122.5  \\  
J030634.26-750133.3  &  M &  14.12 &   88 &  2458035.74364 &     143.2  \\  
J030740.92-610018.8  &  M &  14.80 &   54 &  2458035.54878 &     300.9  \\  
	\hline
	\end{tabular}
	\\ 
	$^1$ F = ESO 2.2m + FEROS; K = Keck + HIRES; M = Magellan + MIKE. \\ 
	$^2$ Signal-to-noise ratio per pixel near 4500\AA. \\ 
	This table is published in its entirety in the electronic edition of the paper. A portion is shown here for guidance regarding its form and content.
\end{table*}

\section{Analysis}

The stellar parameters were derived using the same approach as described in \citet{Norris:2013ab} and \citet{Yong:2013ab} to ensure that the current study is on the same scale and that the two samples can be combined. Effective temperatures (\teff) were from the spectrophotometric fits to the WiFeS spectra but adjusted by +50K for the following reason. In \citet{Norris:2013ab}, \teff\ was the mean from the spectrophotometric flux fitting method, Balmer line profiles and an empirical relation between the H$_\delta$ index HP2 and \teff\ from the infrared flux method \citep{Casagrande:2010aa}, red giants from \citet{Cayrel:2004aa} and from \citet{Norris:2013ab}. In that study, the effective temperatures from the spectrophotometric flux fitting method were, on average, 50K cooler than the mean value. Therefore in the present work, we increase those values by 50K to be on the same scale. 

As in \citet{Yong:2013ab}, surface gravities (\logg) were adopted from the $Y^2$ isochrones \citep{Demarque:2004aa} assuming an age of 13 Gyr and [$\alpha$/Fe] = +0.3. For five objects, the spectrophotometric flux fitting indicated surface gravities that were inconsistent with being red giant branch stars. These five stars are horizontal branch or asymptotic giant branch objects. For the remaining objects, the mean difference in surface gravity (high-resolution analysis $-$ ANU 2.3m) is +0.28 ($\sigma$ = 0.37). We also checked our surface gravities by comparing against values obtained using Gaia EDR3 parallaxes \citep{gaia:edr3} assuming a mass of 0.8 $M_\odot$. For objects with fractional errors in parallax $<$20\%, the difference in \logg\ (this study $-$ Gaia) is $-$0.19 ($\sigma$ = 0.33). We regard the agreement as satisfactory given the 0.3 dex uncertainty in \logg\ values from the spectrophotometric fits \citep{DaCosta:2019aa}. 

\begin{figure*}
	\includegraphics[width=.80\hsize]{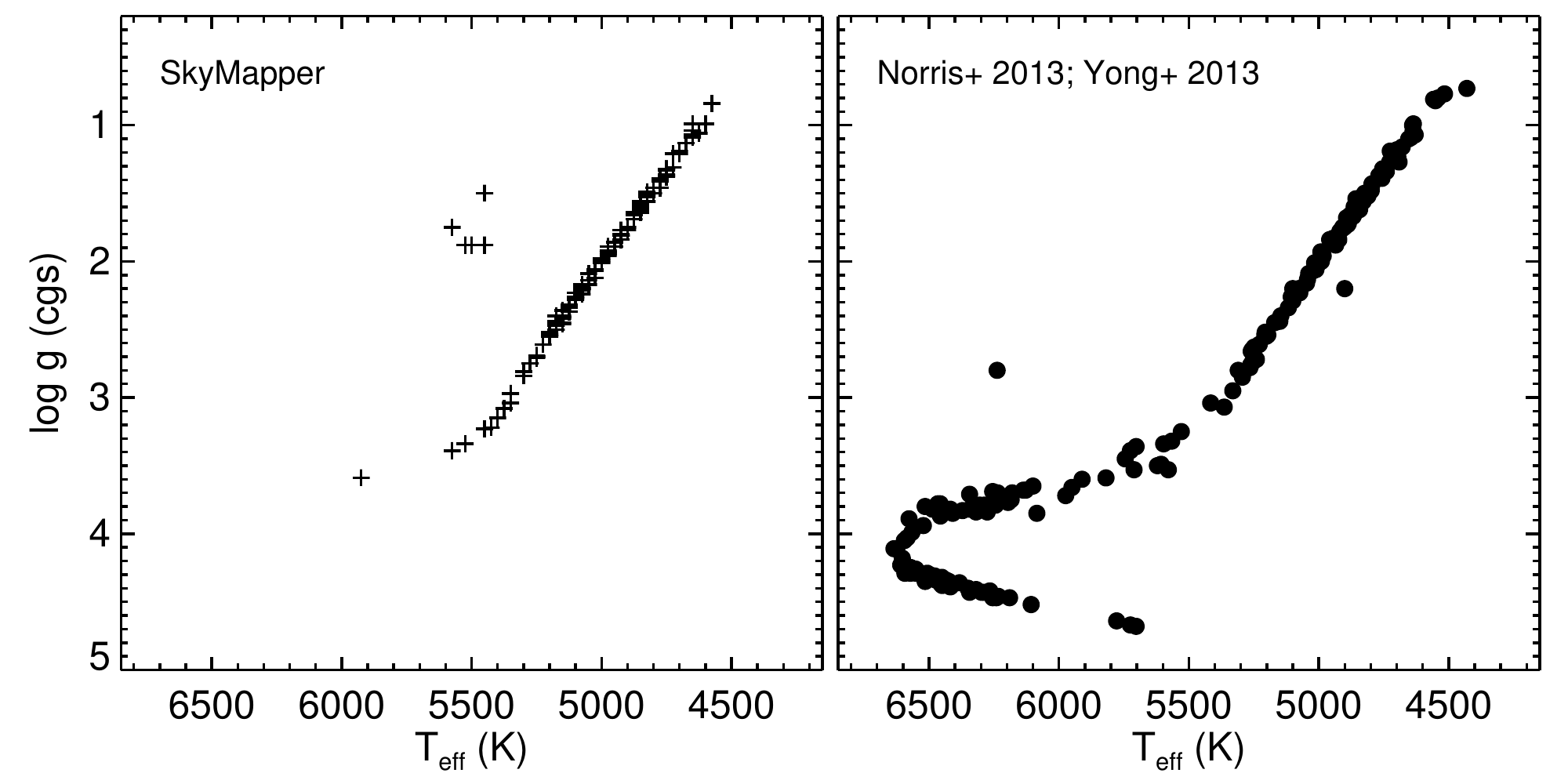}
    \caption{\teff\ vs.\ \logg\ for the current SkyMapper sample (left) and for the \citet{Norris:2013ab} and \citet{Yong:2013ab} sample (right).}
    \label{fig:param}
\end{figure*}

Model atmospheres were taken from the $\alpha$-enhanced, [$\alpha$/Fe] = +0.4, NEWODF grid of ATLAS9 models by \citet{Castelli:2003aa}. These one-dimensional (1D), plane-parallel, local thermodynamic equilibrium (LTE) models were computed using a microturbulent velocity of 2 \kms\ and no convective overshooting. Using software described in \citet{Allende-Prieto:2004aa}, we interpolated within the grid to produce models with the required \teff, \logg\ and [M/H]. In Figure \ref{fig:param} (left panel), we present the location of the program stars in the \teff\ vs.\ \logg\ plane. For comparison, we also present the \citet{Norris:2013ab} sample in the right panel. It is clear that the present sample consists entirely of evolved stars, i.e., objects on the red giant branch, horizontal branch or asymptotic giant branch. 

Equivalent widths were measured for a set of lines in all program stars (see Table \ref{tab:ew}). The line list is identical to the one used in \citet{Norris:2013ab} and \citet{Yong:2013ab}. Radial velocities (see Table \ref{tab:obs}) were measured by comparing the observed and predicted wavelengths of the lines for which equivalent widths were measured. The typical standard deviation was 0.5 \kms. 

Using the LTE stellar line analysis program MOOG \citep{Sneden:1973aa,Sobeck:2011aa}, we computed abundances for \fei\ and \feii\ lines. The microturbulent velocity, \vt, was determined by forcing the abundances from \fei\ lines to have no trend with the reduced equivalent width, log($W_\lambda/\lambda$). The metallicity, [Fe/H], was inferred from \fei\ lines. We recognise that \feii\ lines are less affected by non-LTE effects  \citep{Asplund:2005aa,Bergemann:2012aa,Lind:2012aa,Amarsi:2016aa}. However, there are considerably fewer \feii\ lines in the program stars compared to the number of \fei\ lines, and we were also interested in being consistent with the \citet{Yong:2013ab} study which adopted the same methodology. 

\begin{table}
	\centering
	\caption{Line list and equivalent width measurements.}
	\label{tab:ew}
	\begin{tabular}{lccrrr}
	\hline
	Wavelength &
	Species &
	L.E.P. &
	$\log gf$ &
	EW &
	EW \\
	(\AA) &
	&
	(eV) &
	&
	(m\AA) &
	(m\AA) \\
	\hline
           &       &       &         & 0016-0241 & 0054-8441 \\
   5889.95 &  11.0 &  0.00 &    0.10 &     107.8 &      83.2 \\
   5895.92 &  11.0 &  0.00 & $-$0.19 &      82.2 &      53.0 \\
   3829.36 &  12.0 &  2.71 & $-$0.21 &     117.4 &    \ldots \\
   3832.30 &  12.0 &  2.71 &    0.15 &     152.2 &    \ldots \\
   3838.29 &  12.0 &  2.72 &    0.41 &    \ldots &    \ldots \\
   4571.10 &  12.0 &  0.00 & $-$5.39 &      17.4 &    \ldots \\
   5172.68 &  12.0 &  2.71 & $-$0.38 &     145.5 &     108.1 \\
   5183.60 &  12.0 &  2.72 & $-$0.16 &     151.9 &     118.5 \\
   5528.41 &  12.0 &  4.34 & $-$0.34 &      28.7 &    \ldots \\
   3944.01 &  13.0 &  0.00 & $-$0.64 &     113.5 &      53.9 \\
	\hline
	\end{tabular}
	\\
		This table is published in its entirety in the electronic
edition of the paper. A portion is shown here for guidance regarding its form
and content.
\end{table}

We then compared the derived metallicity, [Fe/H], with the value assumed when generating the model atmosphere, [M/H]. If the difference exceeded 0.2 dex, we computed an updated model atmosphere with [M/H]$_{\rm new}$ = [Fe/H]$_{\rm star}$, and the surface gravity was re-computed (using isochrones as described above but with the updated metallicity). This process was repeated until the stellar parameters converged (usually within an iteration or two). During this process, we removed \fei\ lines that differed from the median abundance by more than 0.5 dex or 3-$\sigma$. Additionally, we were mindful that for the C-rich stars, some lines can be blended with CH so we repeated the entire analysis using a set of lines which we believe are free from CH blending \citep{Norris:2007aa,Norris:2010ab}. Those stars are identified in Table \ref{tab:param1} as "C-rich = 1". Stellar parameters are presented in Table \ref{tab:param1}. 

Recall that some nine stars have S/N $<$ 10. While we present radial velocities and stellar parameters for those objects, we do not present chemical abundances. Additionally, there are nine objects which were observed with multiple telescopes. Seven of these stars were observed with FEROS and MIKE. We present their stellar parameters and radial velocities separately in Tables \ref{tab:obs} and \ref{tab:param1}. For the chemical abundance ratios, however, we only provide measurements from the higher quality MIKE spectra. Two of these stars were observed with HIRES and MIKE. We provide stellar parameters and chemical abundances separately, and use the average values in the figures. 

\begin{table*}
	\caption{Model atmosphere parameters.}
	\label{tab:param1}
	\begin{tabular}{lcccccccccrc}
	\hline
	ID &
	Telescope & 
	\teff &
	\logg &
	\vt &
	[M/H]$_{\rm model}$ &
	[Fe/H]$_{\rm derived}$ &
	C-rich$^1$ &
	CH$^2$ &
	Class \\
	SMSS
	&
	&
	(K) &
	(cgs) &
	(\kms) &
	dex &
	dex &
	&
	&
	\\
	\hline
J001604.23-024105.0     &  M &   5075 &   2.20 &  1.8 & $-$3.1 & $-$3.21 &     0 &     0 &                   \\  
J005420.96-844117.0     &  M &   5275 &   2.75 &  1.3 & $-$3.4 & $-$3.51 &     0 &     0 &                   \\  
J011126.27-495048.4     &  M &   5075 &   2.21 &  1.5 & $-$3.0 & $-$2.94 &     0 &     0 &      Fe-rich      \\  
J020050.19-465735.2     &  M &   5050 &   2.09 &  1.8 & $-$3.7 & $-$3.66 &     0 &     0 &                   \\  
J024246.96-470353.6     &  M &   4775 &   1.41 &  2.1 & $-$3.0 & $-$2.94 &     0 &     0 &                   \\  
J030245.60-281454.0     &  M &   4775 &   1.39 &  2.1 & $-$3.4 & $-$3.50 &     0 &     0 &                   \\  
J030258.53-284326.9     &  M &   5575 &   1.75 &  2.1 & $-$2.7 & $-$2.67 &     0 &     0 &                   \\  
J030428.44-340604.8     &  M &   4750 &   1.32 &  2.2 & $-$3.4 & $-$3.26 &     0 &     0 &         NEMP      \\  
J030634.26-750133.3     &  M &   5075 &   2.20 &  1.6 & $-$3.2 & $-$3.14 &     0 &     0 &                   \\  
J030740.92-610018.8     &  M &   5025 &   2.06 &  1.9 & $-$3.1 & $-$3.09 &     0 &     0 &                   \\  
J030853.27-700140.1     &  M &   4775 &   1.40 &  2.1 & $-$3.2 & $-$3.16 &     1 &     1 &    CEMP           \\  
	\hline
	\end{tabular}
	\\ 
	$^1$ 0 = C-normal  1 = CEMP object  adopting the \citet{Aoki:2007aa} definition and 2 = C-rich when including the \citet{Placco:2014aa}  corrections for the effect of evolutionary status on carbon abundances. \\
	$^2$ 0 = G band maximum depth is $\ge$ 0.75 relative to the continuum and 1 = G band maximum depth in the best fitting synthesis near 4323\AA\ is $<$ 0.75 relative to the continuum. \\
	$^3$ S/N $<$ 10. We report \teff, \logg\ and [Fe/H], but no chemical abundance ratios. \\
	This table is published in its entirety in the electronic edition of the paper. A portion is shown here for guidance regarding its form and content.
\end{table*}

The spectra of a number of cooler stars, despite being `C-normal', i.e., not enhanced in C ([C/Fe] $<$ 0.7), nevertheless show numerous strong CH lines that could potentially \color{black} blend and contaminate the atomic lines. To identify those stars, we utilised the spectrum synthesis of the CH G band near 4300 \AA\ as described later in this section. We identified a CH feature near 4323\AA\ and selected a threshold depth of 0.75 relative to the continuum. For any star in which the best fitting synthetic spectrum reached a depth greater than 0.75, we analysed that object using the CH clean line list. Those stars are flagged in Table \ref{tab:param1} with "G band strength = 1", and there are 21 such objects in the sample. The final stellar parameters are presented in Table \ref{tab:param1} where for the stars which are C-rich or have strong G band strengths, we adopted the results from the CH-clean line list. 

\begin{figure}
	\includegraphics[width=.98\hsize]{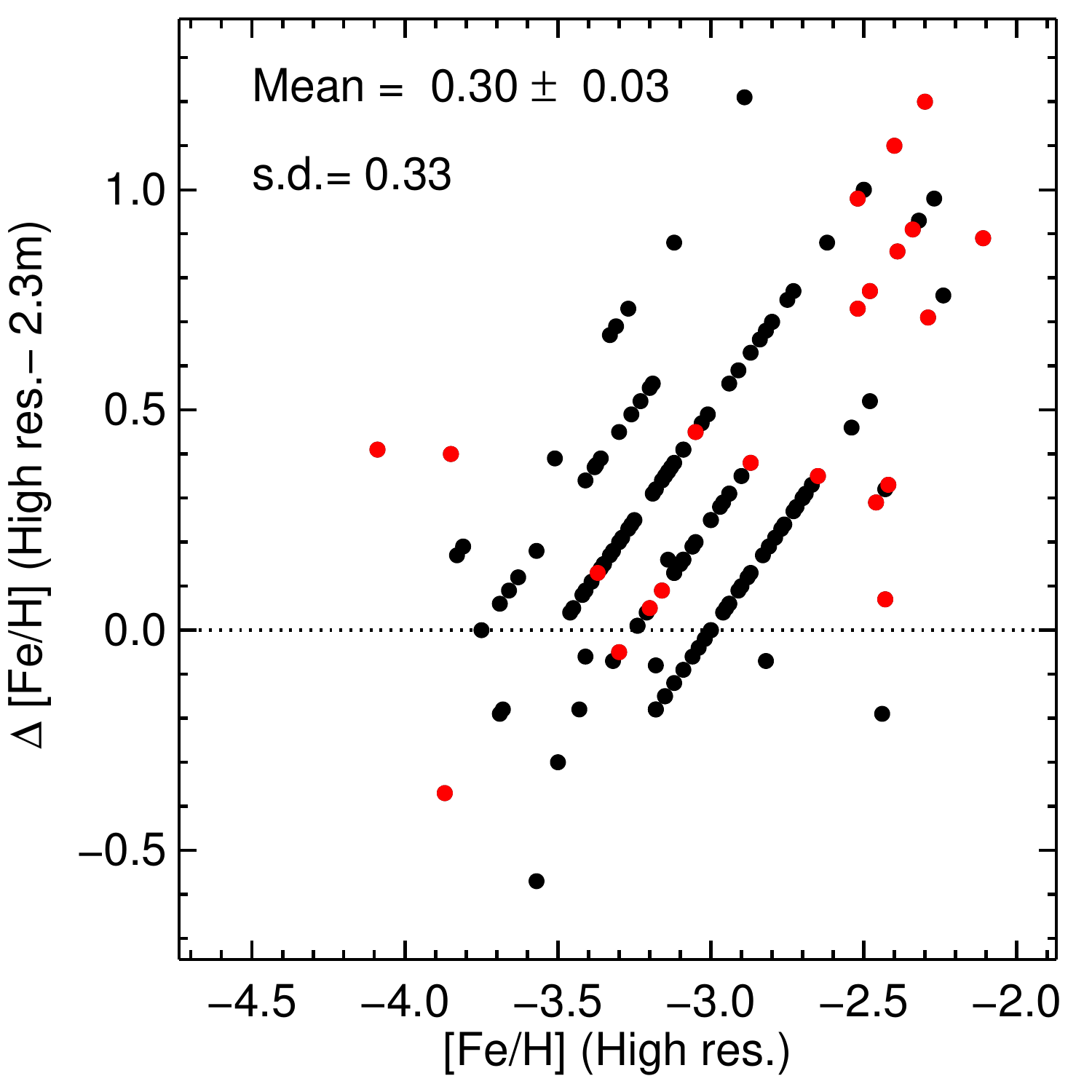}
    \caption{Comparison of the metallicities from the high-resolution Magellan spectra and the intermediate resolution ANU 2.3m spectra. The red symbols are C-rich or those flagged with G band strengths = 1. (Note that the 2.3m metallicities are quantised at the 0.25 dex level, although in some cases multiple observations were averaged.)}
    \label{fig:mdf10.2}
\end{figure}

In Figure \ref{fig:mdf10.2}, we compare the metallicities from the spectrophotometric flux fitting method using the WiFeS spectra and from the analysis of the high dispersion spectra. The metallicities from the high dispersion spectra are, on average, 0.30 $\pm$ 0.03 ($\sigma$ = 0.33) dex higher than the values from the WiFeS spectra. While this difference is slightly larger than the value reported in \citet{DaCosta:2019aa} of 0.04 $\pm$ 0.07 ($\sigma$ = 0.38) dex, we reiterate that the metallicities from the WiFeS spectra are quantised at the 0.25 dex level (in some cases multiple observations were averaged). While we increased \teff\ by 50K, the impact upon [Fe/H] is only +0.03 dex and cannot explain the +0.30 dex difference in metallicity.  

Element abundances were determined from the measured equivalent widths using MOOG. Lines of \scii, \mni, \coi\ and \baii\ are affected by hyperfine splitting and we used data from \citet{Kurucz:1995aa} in our analysis. \baii\ lines are also affected by isotopic splitting and we assumed the $r$-process isotopic composition and hyperfine splitting from \citet{McWilliam:1998aa}. For \euii\ lines we also included isotopic and hyperfine splitting from \citet{Lawler:2001ab}. 

\begin{figure}
	\includegraphics[width=.89\hsize]{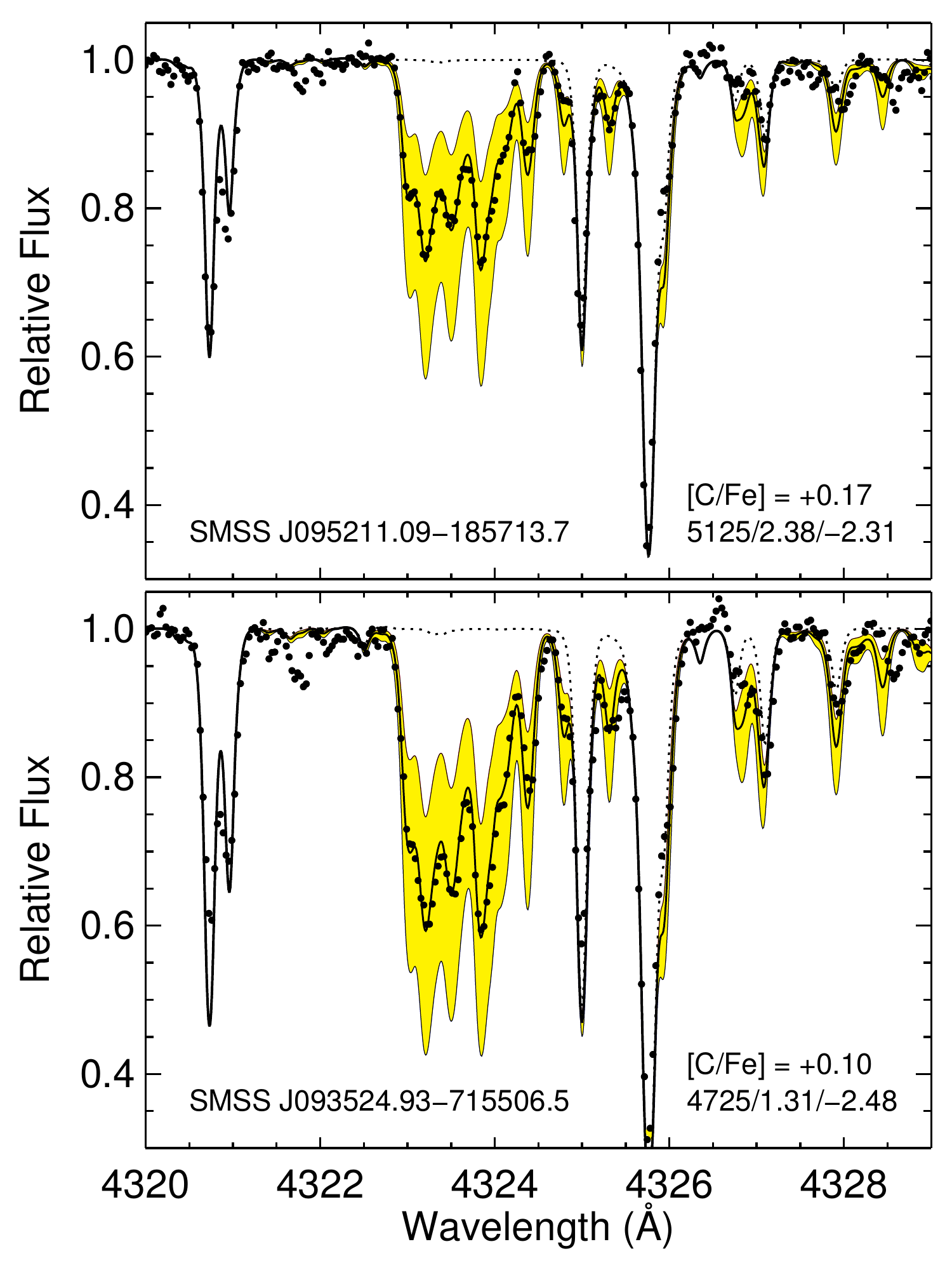}
    \caption{Comparison of observed (filled dots) and synthetic spectra in the region near 4325\AA. Synthetic spectra with no C, [C/Fe]= $-$9, are shown as thin dotted lines. The best-fitting synthetic spectra are the thick black lines and the yellow shaded regions indicate $\pm$0.3 dex from the best fit. The stellar parameters \teff/\logg/[Fe/H] are shown.}
    \label{fig:c_spec}
\end{figure}

\begin{figure}
	\includegraphics[width=.89\hsize]{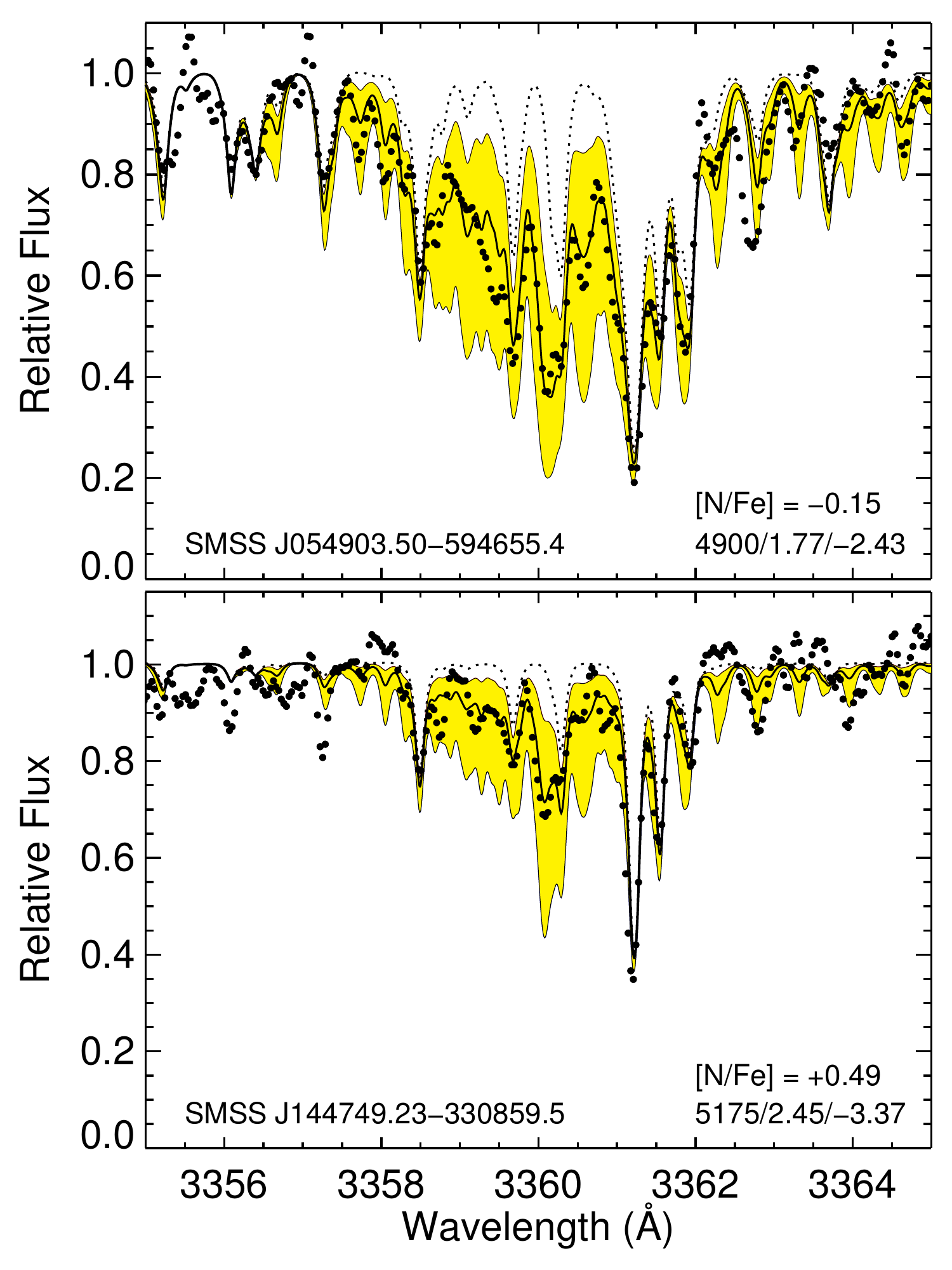}
    \caption{Same as Figure \ref{fig:c_spec} but for [N/Fe] in the region 3355\AA\ to 3365\AA. The yellow shaded regions indicate $\pm$0.5 dex from the best fit.}
    \label{fig:n_spec}
\end{figure}

For Li, C and N, we measured abundances via spectrum synthesis of the 6707.8\AA\ \lii\ line, the (0–0) and (1–1) bands of the $A-X$ electronic transitions of the CH molecule (4290\AA\ to 4330\AA) and the NH molecule (3350\AA\ to 3370\AA), respectively. We computed synthetic spectra using MOOG and adjusted the abundance until the observed and synthetic spectra were in agreement (see Figures \ref{fig:c_spec} and \ref{fig:n_spec}). The broadening was determined using a Gaussian which represents the combined effects of the instrumental profile, atmospheric turbulence and stellar rotation. The typical uncertainties in the A(Li), [C/Fe] and [N/Fe] abundances are 0.2, 0.3 and 0.4 dex, respectively. 

For a subset of objects we used spectrum synthesis to measure abundances for the following species: \zni\ (4722.16, 4810.53\AA), \yii\ (4398.01, 4883.68, 4900.12\AA), \zrii\ (3998.96, 4149.20, 4156.27, 4208.98\AA) and/or \euii\ (4129.72, 4205.04\AA). The typical uncertainties in the [X/Fe] ratios for Zn, Y, Zr and Eu are 0.2, 0.2, 0.15 and 0.3 dex, respectively. We note that the numbers of stars in which we could measure Zn, Y, Zr and Eu abundances were 35, 23, 27 and 26, respectively. Given that we were unable to measure these elements in the majority of stars, we defer the discussion of those abundances until Sec 4.4.  

As with the analysis of Fe lines, we were careful to avoid blending from CH lines for C-rich stars or those with strong G band strengths. Therefore, we repeated the analysis of atomic lines using a line list that was unaffected by CH lines and the stellar parameters determined from a similarly CH-free line list. The abundance ratios are presented in Tables \ref{tab:li}, \ref{tab:c}, \ref{tab:n}, \ref{tab:abun} and \ref{tab:znyzreu}, and we adopted the solar abundances from \citet{Asplund:2009aa}. 

\begin{table}
	\centering
	\caption{Lithium abundances (note that this table excludes the majority of objects with no measurements).}
	\label{tab:li}
	\begin{tabular}{lcc}
	\hline
	ID &
	A(Li)$_{\rm LTE}$ &
	A(Li)$_{\rm 3D NLTE^1}$ \\
	\hline
SMSS~J030634.26-750133.3  & 1.01  & 1.03 \\
SMSS~J034504.76-724732.2  & 1.51  & 1.56 \\
SMSS~J043800.94-831932.8  & 0.95  & 0.91 \\
SMSS~J044147.05-484842.9  & 1.12  & 1.14 \\
SMSS~J054903.50-594655.4  & 1.19  & 1.11 \\
SMSS~J062445.32-623003.7  & 1.28  & 1.28 \\
SMSS~J085210.25-761250.2  & 1.40  & 1.43 \\
SMSS~J103235.57-131520.2  & 1.05  & 1.07 \\
SMSS~J121709.12-272103.6  & 1.16  & 1.16 \\
SMSS~J125142.79-424304.4  & 1.13  & 1.15 \\
SMSS~J145536.24-340538.2  & 0.94  & 0.96 \\
SMSS~J154340.00-831819.5  & 1.12  & 1.06 \\
SMSS~J154634.19-081030.9  & 1.14  & 1.17 \\
SMSS~J163040.08-715639.1  & 1.21  & 1.27 \\
SMSS~J165512.00-725554.9  & 0.99  & 0.92 \\
SMSS~J172313.82-602320.6  & 1.00  & 0.98 \\
SMSS~J172604.29-590656.1  & 1.09  & 1.08 \\
SMSS~J212110.47-611758.9  & 0.99  & 1.01 \\
SMSS~J213402.81-622421.1  & 1.13  & 1.13 \\
SMSS~J214716.16-081546.9  & 0.95  & 0.97 \\
SMSS~J215842.28-202915.8  & 0.88  & 0.87 \\
	\hline
	\end{tabular}
	\\
	$^1$ 3D NLTE corrections from \citet{Wang:2021aa}.
\end{table}

\begin{table}
	\centering
	\caption{Carbon abundances and the corrections for evolutionary status
from \citet{Placco:2014aa}.}
	\label{tab:c}
	\begin{tabular}{lrcr}
	\hline
	ID &
	A(C) &
	C$_{\rm corr}$ &
	[C/Fe]$_{\rm corr}$ \\
	&
	(dex) &
	(dex) &
	(dex) \\
	\hline
SMSS~J001604.23-024105.0 &    5.64 &    0.01 &        0.43 \\ 
SMSS~J005420.96-844117.0 & $<$5.86 &    0.01 &     $<$0.95 \\ 
SMSS~J011126.27-495048.4 &    5.74 &    0.01 &        0.26 \\ 
SMSS~J020050.19-465735.2 &    5.33 &    0.01 &        0.57 \\ 
SMSS~J024246.96-470353.6 &    5.40 &    0.52 &        0.43 \\ 
SMSS~J030245.60-281454.0 & $<$4.72 &    0.51 &     $<$0.30 \\ 
SMSS~J030258.53-284326.9 & $<$6.41 &    0.20 &     $<$0.85 \\ 
SMSS~J030428.44-340604.8 &    5.12 &    0.57 &        0.52 \\ 
SMSS~J030634.26-750133.3 &    5.53 &    0.01 &        0.25 \\ 
SMSS~J030740.92-610018.8 &    5.82 &    0.02 &        0.50 \\ 
	\hline
	\end{tabular}
	\\
		This table is published in its entirety in the electronic
edition of the paper. A portion is shown here for guidance regarding its form
and content.
\end{table}

\begin{table}
	\centering
	\caption{Nitrogen abundances (note that this table excludes the
majority of objects with no measurements).}
	\label{tab:n}
	\begin{tabular}{lcrr}
	\hline
	ID &
	A(N) &
	[N/Fe] &
	NEMP$^1$ \\
	&
	(dex) &
	(dex) \\
	\hline
 SMSS~J030428.44-340604.8 &   5.35 &       0.78 &     1 \\ 
 SMSS~J031703.94-374047.2 &   5.45 &       0.89 &     1 \\ 
 SMSS~J052313.34-621822.5 &   5.85 &       0.78 &     1 \\ 
 SMSS~J054650.97-471407.9 &   6.35 &       2.61 &     1 \\ 
 SMSS~J054903.50-594655.4 &   5.25 &    $-$0.15 &     0 \\ 
 SMSS~J054913.80-453904.0 &   4.95 &       0.27 &     0 \\ 
 SMSS~J062445.32-623003.7 &   5.15 &    $-$0.25 &     0 \\ 
 SMSS~J091043.10-144418.5 &   5.55 &       1.13 &     1 \\ 
 SMSS~J095211.09-185713.7 &   4.95 &    $-$0.54 &     0 \\ 
 SMSS~J095246.98-085554.0 &   5.05 &       0.09 &     0 \\ 
 SMSS~J102410.14-082802.8 &   6.05 &       0.54 &     1 \\ 
 SMSS~J103819.28-284817.9 &   5.25 &       1.11 &     1 \\ 
 SMSS~J110901.23+075441.7 &   5.55 &       0.90 &     1 \\ 
 SMSS~J121709.12-272103.6 &   4.95 &    $-$0.36 &     0 \\ 
 SMSS~J144749.23-330859.5 &   4.95 &       0.49 &     0 \\ 
 SMSS~J151044.04-395653.6 &   5.75 &       0.46 &     0 \\ 
 SMSS~J181200.10-463148.8 &   5.55 &       1.02 &     1 \\ 
 SMSS~J185358.63-555400.1 &   5.75 &       0.94 &     1 \\ 
 SMSS~J190836.24-401623.5 &   6.15 &       1.65 &     1 \\ 
 SMSS~J213402.81-622421.1 &   4.55 &    $-$0.16 &     0 \\ 
	\hline
	\end{tabular}
	\\
	$^1$ 1 = NEMP object \citep{Johnson:2007aa} adopting the uncorrected C abundances. \\
\end{table}

\begin{table*}
	\caption{Chemical abundances (Na-Ba) for the program stars.}
	\label{tab:abun}
	\begin{tabular}{lcccrr}
	\hline
	ID &
	A(X) &
	N$_{\rm lines}$ & 
	s.e.$_{\log \epsilon X}$ &
	[X/Fe] &
    Total Error \\
	\hline
NaI   \\ 
SMSS~J001604.23-024105.0     &    3.04 & 2      &    0.08 &       0.00 &    0.11 \\
SMSS~J005420.96-844117.0     &    2.79 & 2      &    0.12 &       0.07 &    0.12 \\
SMSS~J011126.27-495048.4     &    3.10 & 2      &    0.14 &    $-$0.20 &    0.15 \\
SMSS~J020050.19-465735.2     &  \ldots & \ldots &  \ldots &     \ldots &  \ldots \\
SMSS~J024246.96-470353.6     &    4.09 & 1      &  \ldots &       0.79 &    0.17 \\
SMSS~J030245.60-281454.0     &    3.40 & 2      &    0.07 &       0.66 &    0.14 \\
SMSS~J030258.53-284326.9     &    4.11 & 2      &    0.29 &       0.54 &    0.30 \\
SMSS~J030428.44-340604.8     &    2.81 & 2      &    0.07 &    $-$0.17 &    0.11 \\
SMSS~J030634.26-750133.3     &    2.91 & 2      &    0.06 &    $-$0.19 &    0.11 \\
SMSS~J030740.92-610018.8     &    3.04 & 2      &    0.13 &    $-$0.11 &    0.13 \\
	\hline
	\end{tabular}
	\\
		This table is published in its entirety in the electronic
edition of the paper. A portion is shown here for guidance regarding its form
and content.
\end{table*}

\begin{table*}
	\centering
	\caption{Chemical abundances for Zn, Y, Zr and Eu (note that this table excludes the
majority of objects with no measurements).}
	\label{tab:znyzreu}
\begin{tabular}{lcccrr}
	\hline
	ID &
	A(X) &
	N$_{\rm lines}$ & 
	s.e.$_{\log \epsilon X}$ &
	[Eu/Fe] &
	Total Error \\
	\hline
Zn I \\
SMSS~J034749.80-751351.7 &   2.08 &   1 &    \ldots &      0.02 &   0.30 \\
SMSS~J043800.94-831932.8 &   1.94 &   1 &    \ldots &      0.38 &   0.30 \\
SMSS~J050247.62-642915.9 &   2.45 &   2 &      0.10 &      0.35 &   0.21 \\
SMSS~J052313.34-621822.5 &   2.19 &   1 &    \ldots &      0.39 &   0.30 \\
SMSS~J054903.50-594655.4 &   2.10 &   2 &      0.09 &   $-$0.03 &   0.21 \\
SMSS~J054913.80-453904.0 &   1.85 &   2 &      0.06 &      0.45 &   0.21 \\
SMSS~J062445.32-623003.7 &   2.04 &   1 &    \ldots &   $-$0.09 &   0.30 \\
SMSS~J072146.02-835759.7 &   2.44 &   2 &      0.04 &   $-$0.02 &   0.21 \\
SMSS~J084327.83-141513.3 &   2.68 &   2 &      0.03 &      1.41 &   0.21 \\
SMSS~J091117.11-264637.1 &   2.39 &   2 &      0.02 &      0.10 &   0.21 \\
	\hline
	\end{tabular}
	\\
			This table is published in its entirety in the
electronic edition of the paper. A portion is shown here for guidance regarding
its form and content.
\end{table*}

The chemical abundances are affected by uncertainties in the model atmospheres and we estimated those values to be \teff\ $\pm$ 100K, \logg\ $\pm$ 0.3 dex, \vt\ $\pm$ 0.3 \kms\ and [M/H] $\pm$ 0.3 dex. We repeated the analysis varying the stellar parameters, one at a time, assuming that the errors are symmetric for positive and negative values. We present those uncertainties in Table \ref{tab:abun_err} in which the final column is the accumulated error in which the four values are added in quadrature. To obtain the total error presented in Table \ref{tab:abun}, we update the random error in that table (s.e.$_{\log \epsilon}$\footnote{Standard error of the mean}) by max(s.e.$_{\log \epsilon}$,0.20/$\sqrt{(N_{\rm lines})}$ where the second term is what would be expected for a set of $N_{\rm lines}$ with a dispersion of 0.20 dex (a conservative estimate for the abundance dispersion based on \fei\ lines). The total error is obtained by adding this updated random error in quadrature with the error from the stellar parameters presented in Table \ref{tab:abun_err}. 

\begin{table*}
	\caption{Abundance errors from uncertainties in atmospheric parameters.}
	\label{tab:abun_err}
	\begin{tabular}{llrrrrc}
	\hline
	ID &
	Species &
	$\Delta$ \teff & 
	$\Delta$ \logg &
	$\Delta$ \vt &
	$\Delta$ [M/H] &
	$\Delta$ [X/Fe] \\
	&
	&
	(+100 K) &
	(+0.3 dex) &
	(+0.3 \kms) &
	(+0.3 dex) \\
	\hline
SMSS~J001604.23-024105.0      & $\Delta$[FeI/H]    &    0.06 & $-$0.02 & $-$0.09 &    0.01 &    0.11 \\
SMSS~J001604.23-024105.0      & $\Delta$[FeII/H]   &    0.00 &    0.10 & $-$0.02 &    0.00 &    0.10 \\
SMSS~J001604.23-024105.0      & $\Delta$[Fe/H]     &    0.06 & $-$0.01 & $-$0.09 &    0.01 &    0.10 \\
SMSS~J001604.23-024105.0      & $\Delta$[NaI/Fe]   & $-$0.01 & $-$0.02 & $-$0.01 & $-$0.01 &    0.02 \\
SMSS~J001604.23-024105.0      & $\Delta$[MgI/Fe]   &    0.00 & $-$0.05 & $-$0.00 &    0.00 &    0.05 \\
SMSS~J001604.23-024105.0      & $\Delta$[AlI/Fe]   & $-$0.01 & $-$0.02 & $-$0.05 & $-$0.01 &    0.06 \\
SMSS~J001604.23-024105.0      & $\Delta$[SiI/Fe]   &    0.01 & $-$0.08 & $-$0.04 &    0.00 &    0.09 \\
SMSS~J001604.23-024105.0      & $\Delta$[CaI/Fe]   & $-$0.02 &    0.00 &    0.05 &    0.00 &    0.05 \\
SMSS~J001604.23-024105.0      & $\Delta$[ScII/Fe]  & $-$0.03 &    0.10 & $-$0.02 & $-$0.01 &    0.11 \\
SMSS~J001604.23-024105.0      & $\Delta$[TiI/Fe]   &    0.00 &    0.00 &    0.07 &    0.00 &    0.07 \\
SMSS~J001604.23-024105.0      & $\Delta$[TiII/Fe]  & $-$0.03 &    0.09 & $-$0.01 & $-$0.01 &    0.10 \\
SMSS~J001604.23-024105.0      & $\Delta$[CrI/Fe]   &    0.00 &    0.00 &    0.03 & $-$0.01 &    0.03 \\
SMSS~J001604.23-024105.0      & $\Delta$[MnI/Fe]   &    0.01 & $-$0.01 & $-$0.02 & $-$0.01 &    0.03 \\
SMSS~J001604.23-024105.0      & $\Delta$[CoI/Fe]   &    0.00 &    0.00 & $-$0.00 & $-$0.01 &    0.01 \\
SMSS~J001604.23-024105.0      & $\Delta$[NiI/Fe]   &    0.01 &    0.00 & $-$0.03 &    0.00 &    0.04 \\
SMSS~J001604.23-024105.0      & $\Delta$[SrII/Fe]  & $-$0.02 &    0.09 & $-$0.10 &    0.00 &    0.14 \\
SMSS~J001604.23-024105.0      & $\Delta$[BaII/Fe]  & $-$0.02 &    0.11 &    0.07 &    0.00 &    0.13 \\
	\hline
	\end{tabular}
	\\
		This table is published in its entirety in the electronic
edition of the paper. A portion is shown here for guidance regarding its form
and content.
\end{table*}

\begin{figure*}
	\includegraphics[width=.80\hsize]{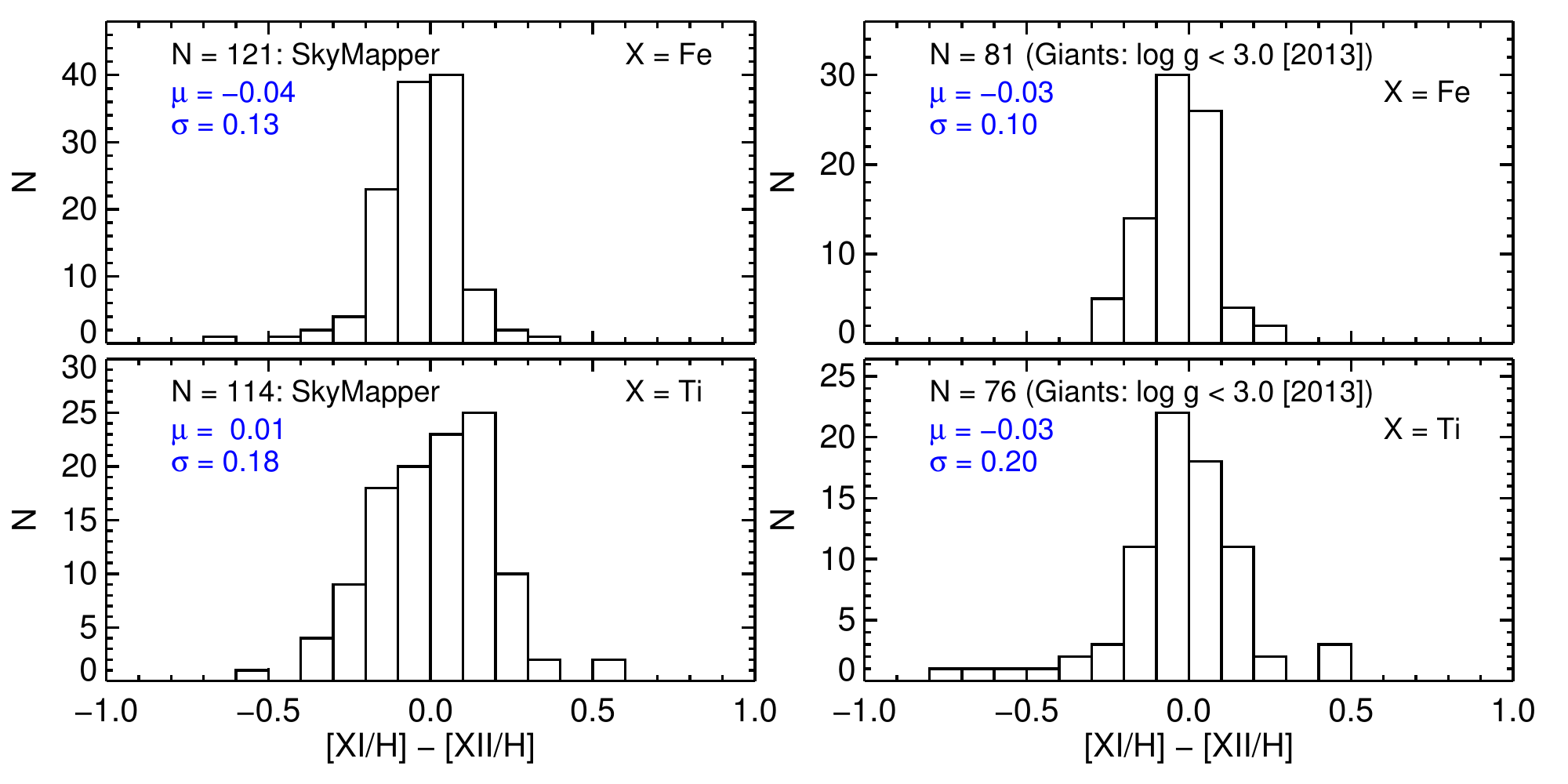}
    \caption{Histograms showing the abundance difference [\fei/H] $-$ [\feii/H] (upper) and [\tii/H] $-$ [\tiii/H] (lower) for $N \ge$ 2 lines and [Fe/H] $\le$ $-$2.5. The current SkyMapper sample is shown in the left panels and giants (\logg\ $<$ 3.0) from the \citet{Norris:2013ab} and \citet{Yong:2013ab} sample in the right panels. In each panel we include the number of stars, mean ($\mu$) and standard deviation ($\sigma$).} 
    \label{fig:fe_ti}
\end{figure*}

For Fe and Ti, abundances have been obtained for the neutral and ionised species. By comparing those abundance ratios, any discrepancy could be attributed to non-LTE effects and/or errors in the surface gravity. For metal-poor stars, the LTE abundances obtained from neutral species are expected to be underestimated due to overionisation \citep{Thevenin:1999aa,Mashonkina:2011aa,Bergemann:2012aa,Lind:2012aa}. That is, we would expect that neglecting non-LTE corrections might lead to negative values of [\fei/H] $-$ [\feii/H] and [\tii/H] $-$ [\tiii/H]. In Figure \ref{fig:fe_ti}, we present the differences between the abundances from \fei\ and \feii\ lines (upper left panel) and from \tii\ and \tiii\ lines (lower left panel). We only consider stars with [Fe/H] $\le$ $-$2.5 and that have two or more lines measured for a given species. For comparison, we show giants from the \citet{Norris:2013ab} and \citet{Yong:2013ab} sample in the right panels. 
In all cases, the histograms are centred near zero. For Fe and Ti, while the abundances from neutral and ionised species are in good agreement, this does not imply that non-LTE effects can be neglected. 

Four of the stars have been previously analysed. The objects are: SMSS~J054913.80-453904.0 (= HE~0547-4539) from \citet{Barklem:2005aa}, 
SMSS~J143511.34-420326.4 (= SMSS~J1435-4203) from \citet{Jacobson:2015aa}, 
SMSS~J030428.44-340604.8 (= HE~0302-3417A) from \citet{Hollek:2011aa} and 
SMSS~J232121.57-160505.4 (= HE~2318-1621) from \citet{Placco:2014ab}.

Different approaches to determining \teff\ and \logg, as well as different line lists and $\log gf$ values result in systematic differences between the previous studies and this work. For example, for stars SMSS~J054913.80-453904.0, SMSS~J143511.34-420326.4, SMSS~J030428.44-340604.8 and SMSS~J232121.57-160505.4 we find [Fe/H] = $-$3.15, $-$2.65, $-$3.26, and $-$3.03, while previous studies report [Fe/H] = $-$3.01, $-$3.15, $-$3.70, and $-$3.67, respectively.  

We will not seek to understand the origin of any of these discrepancies in detail, merely noting that an advantage of our large and homogeneously analysed sample is that within it stars and their abundances can be readily compared with each other. However, we note that the average difference in [X/Fe] for these stars is generally small with the mean difference across 15 elements, in the sense (this work $-$ literature) is $-$0.02 dex with a standard deviation of 0.32 dex.

\section{Results}

\subsection{MDF}

\begin{figure*}
	\includegraphics[width=.80\hsize]{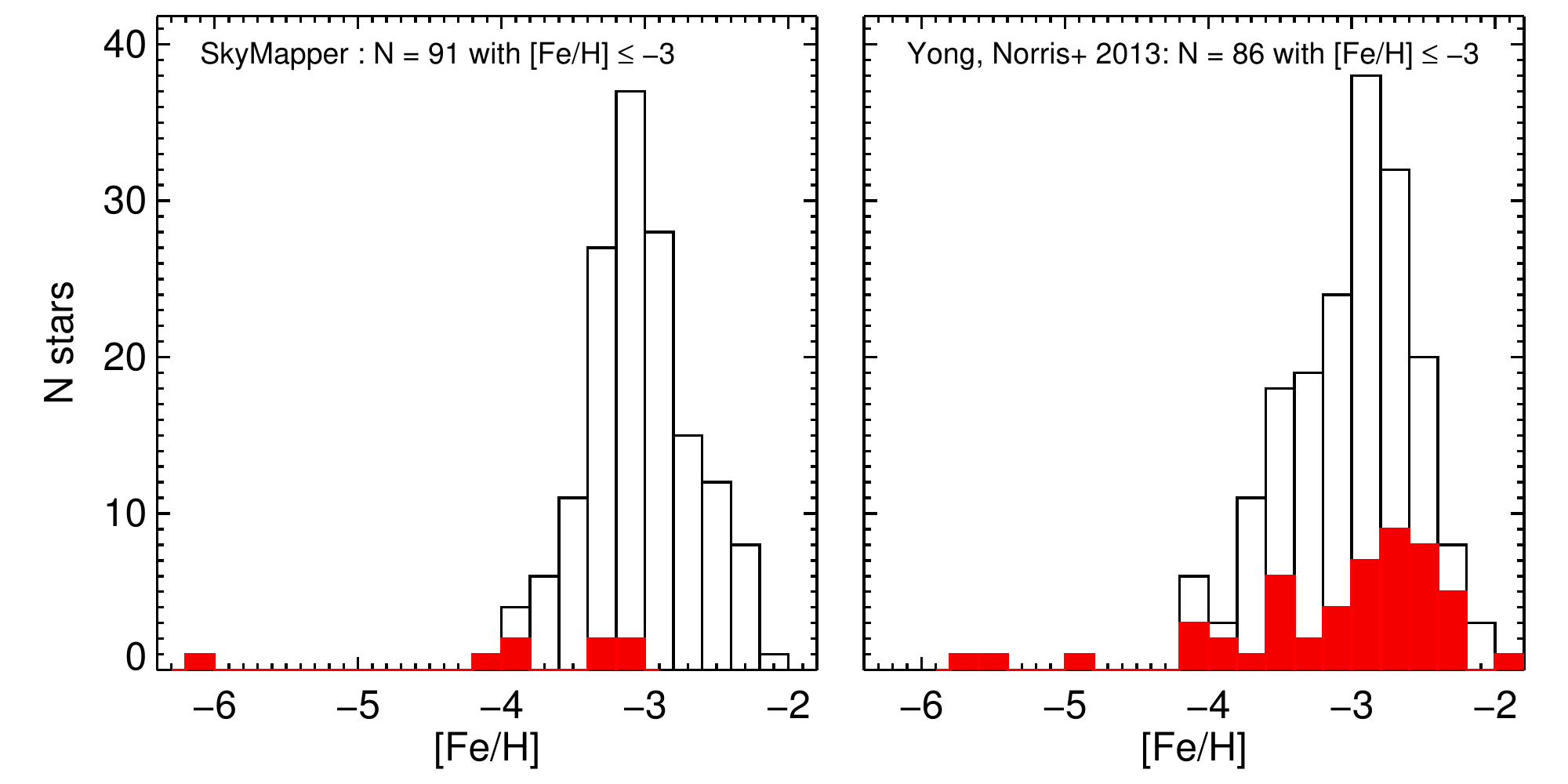}
    \caption{Metallicity distribution function for the current SkyMapper sample including SMSS~J160540.18-144323.1 from \citet{Nordlander:2019aa} (left) and for the \citet{Yong:2013aa} sample (right). The numbers of stars with [Fe/H] $\le$ $-$3 are included in each panel and the red histogram indicates C-rich stars.}
    \label{fig:mdf0}
\end{figure*}

The metallicity distribution function, MDF (usually based on Fe), is a crucial diagnostic tool for understanding low mass star formation in the early universe. In Figure \ref{fig:mdf0}, we plot the MDF for the current sample (left panels) and the \citet{Norris:2013ab} and \citet{Yong:2013ab} sample (right panels). In this figure, we also include SMSS~J160540.18-144323.1 from \citet{Nordlander:2019aa} with [Fe/H] = $-$6.2 which was discovered using SkyMapper DR 1.1 photometry. Unless explicitly mentioned otherwise, in the subsequent discussion we will not include this object. It is 2 dex more iron-poor than the next most iron-poor star in the sample, the observations and analysis differ from the approach in this study, and only a handful of elements have abundance measurements. \citet{Nordlander:2019aa} have already presented a comprehensive analysis and interpretation of this star. There are 91 stars with [Fe/H] $\le$ $-$3.0 in the current sample of which 87 are newly reported objects; the other four known stars are SMSS~J160540.18-144323.1 from \citet{Nordlander:2019aa}, SMSS~J054913.80-453904.0 from \citet{Barklem:2005aa}, SMSS~J030428.44-340604.8 from \citet{Hollek:2011aa}, and SMSS~J232121.57-160505.4 from \citet{Placco:2014ab}. For consistency we have used our [Fe/H] determinations for the latter three stars: [Fe/H] = $-$3.15, $-$3.26 and $-$3.03, respectively. We emphasise that the two samples, this study and \citet{Norris:2013ab}, are completely independent as the 2.3m follow-up observations of SkyMapper candidates has deliberately attempted to exclude previously known stars. 

In order to combine the two samples, there are several considerations that need to be taken into account. For the \citet{Norris:2013ab} sample, the selection biases are described in \citet{Yong:2013aa} (see their Figure 1). There were two main factors to consider. First, the Hamburg ESO Survey (HES) from which most targets were drawn is complete below [Fe/H] = $-$3.0 \citep{Schorck:2009aa,Li:2010aa}. Second, the ratio of HES candidates observed at high resolution relative to the total number of HES candidates as a function of metallicity needs to be included. That ``completeness function'' was presented in \citet{Norris:2013ab}. 

\begin{figure*}
	\includegraphics[width=.80\hsize]{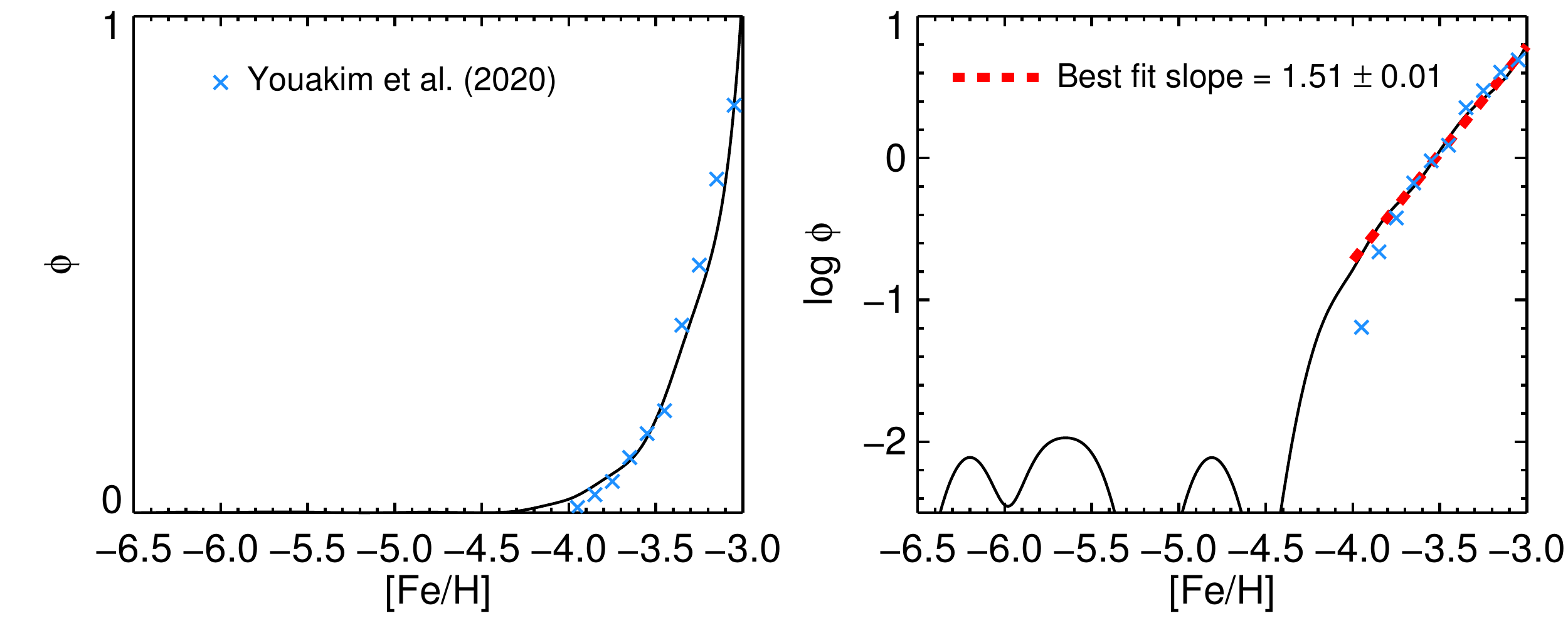}
    \caption{The metallicity distribution function in generalised histogram form (black lines) using linear (left) and logarithmic (right) scales for the combined sample (this study including SMSS~J160540.18-144323.1 from \citet{Nordlander:2019aa} plus \citealt{Norris:2013ab}). In the right panel, we fit the slope between $-$4 $\le$ [Fe/H] $\le$ $-$3. In both panels, we overplot data from \citet{Youakim:2020aa} (corrected with their Gaussian mixture model and colour cuts and normalised at [Fe/H] = $-$3.05) as blue crosses. }
    \label{fig:mdf10.9}
\end{figure*}

For the current sample, candidates were first selected from the SkyMapper metallicity sensitive diagram, $m_i = (v - g)_0 - 1.5 (g-i)_0$ vs.\ $(g-i)_0$, as described in \citet{DaCosta:2019aa}. In principle, more metal-poor objects should have more negative values of $m_i$. However, \citet{DaCosta:2019aa} showed that for objects with [Fe/H] $\le$ $-$2.0, there was little correlation between the metallicity and the $m_i$ value. That is, while the SkyMapper DR1.1 photometry is highly efficient at identifying stars with [Fe/H] $<$ $-$2, it cannot readily discriminate between stars with [Fe/H] = $-$4 and [Fe/H] = $-$2 (the differences in the $m_i$ values are smaller than the typical photometric errors). They also noted that large carbon enhancements can affect the $m_i$ index making CEMP objects appear to be more metal rich. Below [Fe/H] = $-$4, \citet{DaCosta:2019aa} suggest that C-rich stars do not fall outside of the photometric selection window. At [Fe/H] = $-$3.0, they suggest that is likely that some strongly C enhanced objects will fall out of the selection box, although it is a ``complex function of effective temperature, [Fe/H] and [C/Fe], as well as of [N/Fe] and [O/Fe]'' \citep{DaCosta:2019aa}. Above about [Fe/H] = $-$3.5, the bias against C enhanced stars is clearly visible in Figure \ref{fig:mdf10.9} where there is a lack of C rich stars in the current sample (left panel) when compared to the previous sample (right panel). 

In the absence of further information (i.e., the SkyMapper selection bias, the [C/Fe] distribution as a function of \teff, \logg\ and [Fe/H], photometric uncertainties etc.), we will cautiously proceed by producing a generalised histogram for the current sample and another for the ``completeness function'' corrected sample from \citet{Norris:2013ab}. That is, each data point is replaced by a unit Gaussian of width 0.15 dex. The Gaussians are summed to produce a realistically smoothed histogram. We normalise both histograms by the numbers of stars in each sample then combine the two distributions. The MDF for the combined sample (including SMSS~J160540.18-144323.1 from \citealt{Nordlander:2019aa}) is presented in Figure \ref{fig:mdf10.9} with linear (left) and logarithmic (right) scales. 

For stars in the range $-$4 $\le$ [Fe/H] $\le$ $-$3, we fit the data using a linear function (right panel) and find a power-law slope of $\Delta$(log N)/$\Delta$[Fe/H] = 1.51 $\pm$ 0.01 dex per dex. This slope is in excellent agreement with the value of 1.5\footnote{The slope was determined from stars in the range $-$4.0 $\le$ [Fe/H] $\le$ $-$2.75.} $\pm$ 0.1 in \citet{DaCosta:2019aa}, but considerably steeper than the canonical 1.0 from the \citet{Hartwick:1976aa} simple model. While we have yet to properly account for the impact of C-rich objects in the SkyMapper selection, our results reinforce how difficult it is to find stars more metal-poor than [Fe/H] = $-$3.0. 

In creating this MDF from the combined sample, there are 177 stars (including  SMSS~J160540.18-144323.1) with [Fe/H] $\le$ $-$3.0. We can compare the fractional uncertainty in the MDF from the \citet{Norris:2013ab} sample with the combined sample. That fractional uncertainty was obtained using the same approach described in \citet{Yong:2013aa}. That is, using Monte Carlo simulations we replaced each data point with a random number drawn from a normal distribution of width 0.15 dex centred at the [Fe/H] of each star. We repeated this process for the entire sample and created a new generalised histogram. For 10,000 new random samples, we produced a generalised histogram for each random sample. Thus, at a given [Fe/H], we have a distribution of 10,000 values (one for each MDF), and we measured the FWHM of that distribution. That FWHM was taken as our estimate of the uncertainty in the MDF. For all values between $-$4 $\le$ [Fe/H] $\le$ $-$3, the fractional uncertainty in the MDF has improved with respect to the analysis in \citet{Yong:2013aa}. At [Fe/H] = $-$4.0, the fractional uncertainty decreased by about 10\%. The greatest improvement was a $\sim$70\% decrease in the fractional uncertainty near [Fe/H] = $-$3.3. 

In Figure \ref{fig:mdf10.9}, the formal uncertainty on the slope is only 0.01. In the Monte Carlo approach described above, we fit each new MDF between $-$4 $\le$ [Fe/H] $\le$ $-$3 and find an average power-law slope of $\Delta$(log N)/$\Delta$[Fe/H] = 1.45 dex per dex with a standard deviation of 0.07 dex. We regard 0.07 dex as a more realistic estimate of the uncertainty in the MDF slope. 

If we generate a MDF using the same approach described above but excluding the C-rich stars, the slope between $-$4 $<$ [Fe/H] $<$ $-$3 is 1.74 $\pm$ 0.02. There are no C-normal stars substantially below [Fe/H] = $-$4, although we do not include the \citet{Caffau:2011aa} star (which is a dwarf and the SkyMapper DR 1.1 sample is dominated by giants). In Figure \ref{fig:mdf0}, C-rich stars are indicated by the red histogram and the MDF for those objects is considerably flatter than for the full sample. 

\citet{Youakim:2020aa} presented the MDF of the Pristine survey based on a photometric sample of 80,000 main-sequence turn-off stars representative of the inner halo of the Galaxy (we overplot their data, corrected with their Gaussian mixture model and colour cuts and normalised at [Fe/H] = $-$3.05, in Figure \ref{fig:mdf10.9}). Overall, they note that the MDF is not well represented by a single power-law but in the metallicity range $-$3.4 $<$ [Fe/H] $<$ $-$2.5, they find a slope of $\Delta$(log N)/$\Delta$[Fe/H] = +1.0 $\pm$ 0.1. While we would like to also examine our MDF over the same metallicity interval as \citet{Youakim:2020aa}, our sample is incomplete above [Fe/H] $\simeq$ $-$2.6. That is, when generating an MDF using our combined sample, there is an artificial turnover near [Fe/H] $\simeq$ $-$2.6. Therefore, we consider a slightly different metallicity range $-$3.4 $<$ [Fe/H] $<$ $-$2.7, in which we find that our MDF has a slope of +1.07 $\pm$ 0.04 which is in excellent agreement with the \citet{Youakim:2020aa} value. 

Similarly, based on a large sample of candidate metal-poor giants selected from SkyMapper DR2 photometry \citep{Onken2019:aa}, \citet{Chiti2021:aa} find that the MDF in the metallicity range $-$3.0 $\leq$ [Fe/H] $\leq$ $-$2.3 is well-fit by a power-law with a slope of 
$\Delta$(log N)/$\Delta$[Fe/H] = +1.53 $\pm$ 0.10.

In the metallicity range $-$4.0 $<$ [Fe/H] $<$ $-$3.4, we find the MDF has a slope of 1.59 $\pm$ 0.02. Excluding C-rich stars, the slope increases to 2.11 $\pm$ 0.05. Again this is not dissimilar to the results of \citet{Youakim:2020aa} who find a slope of +2.0 $\pm$ 0.2 for stars in their sample with [Fe/H] $<$ $-$3.5 dex. Both studies (i.e., \citeauthor{Youakim:2020aa} and this work) agree on a significant change of slope somewhere around [Fe/H] = $-$3.5 to $-$4.

\subsection{Li, C and N}

Lithium abundances, A(Li), were measured in 21 stars and are presented in Table \ref{tab:li}. We include LTE and NLTE abundances where that latter makes use of corrections from \citet{Wang:2021aa}. None of the program stars are enhanced in lithium. 

\begin{figure*}
	\includegraphics[width=.80\hsize]{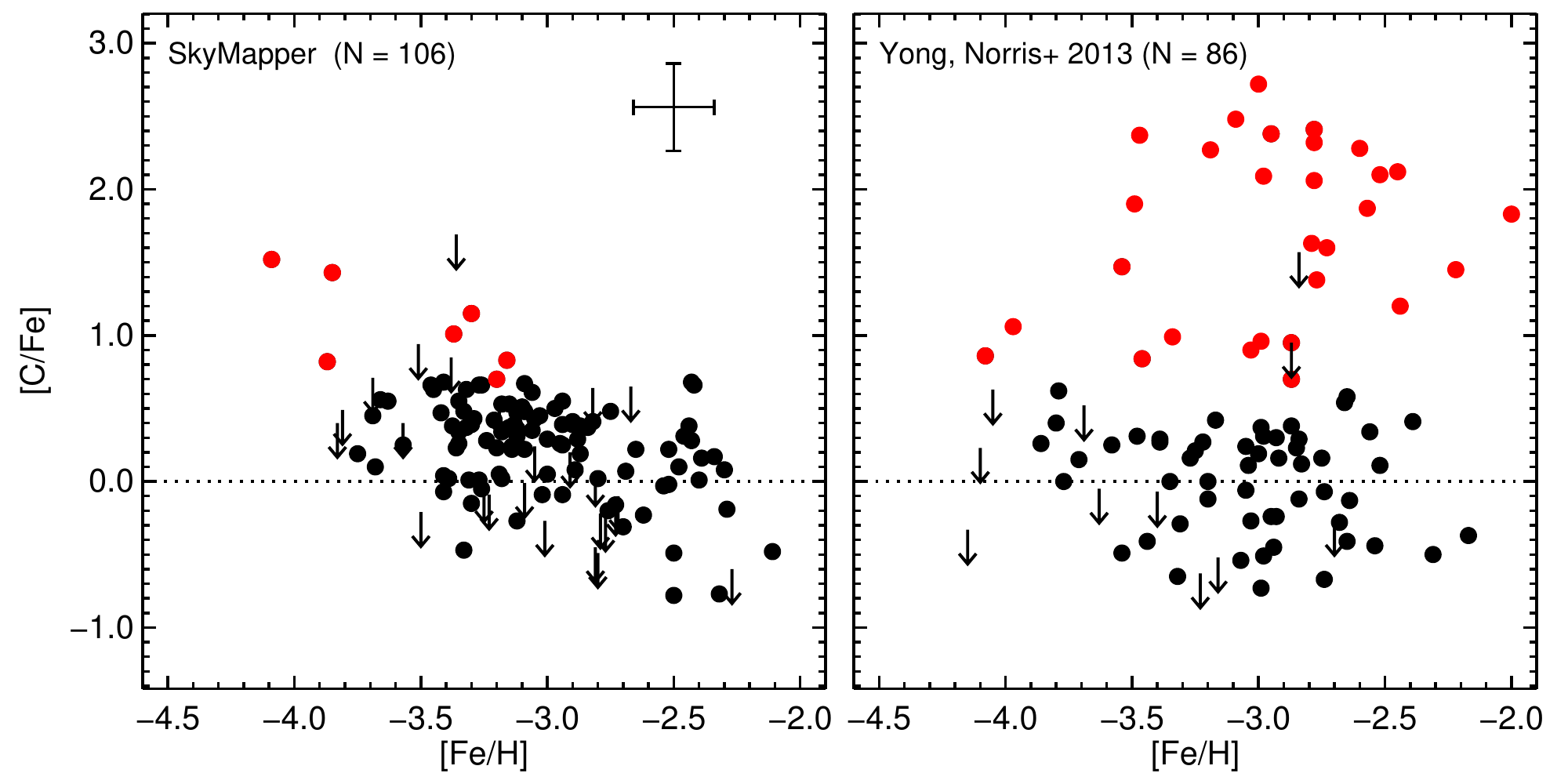}
    \caption{[C/Fe] vs.\ [Fe/H] for the current SkyMapper sample (left) and giants (\logg\ $\le$ 3.0) from the \citet{Yong:2013aa} sample (right). The red symbols are C-rich objects and a representative error bar is included in the top right corner (left panel). (All C abundances are ``observed'', i.e., without correction for evolutionary status.)}
    \label{fig:C}
\end{figure*}

Carbon abundance ratios are presented in Table \ref{tab:c} in which we list the evolutionary corrections from \citet{Placco:2014aa}. We note in particular, that in order to enable a comparison with our previous work, we will utilise the uncorrected carbon abundances unless explicitly stated otherwise. In Figure \ref{fig:C} we plot [C/Fe] versus [Fe/H] for the current sample (left) and the \citet{Norris:2013ab} and \citet{Yong:2013ab} sample (right). For the C-normal populations, the two samples exhibit similar behaviour. For the C-rich population, however, it is clear that the current sample lacks CEMP stars (only seven\footnote{The seven CEMP stars include six CEMP-no objects. The remaining object, SMSS~J030853.27-700140.1, has [Ba/Fe] = +0.09 which does not allow it to be placed into any of the established sub-classes: CEMP-r, CEMP-s, CEMP-r/s and CEMP-no. \color{black} } are present plus a further 13 when taking into account evolutionary corrections\footnote{Those stars are identified in Table \ref{tab:param1} as "C-rich = 2". }) as well as exhibiting a lack of stars with [C/Fe] substantially above +1.5. In contrast, in the right-hand panels of Figure \ref{fig:C} there are some 31 CEMP stars with [C/Fe] $\ge$ +0.7. The lack of very C enhanced stars among samples selected from SkyMapper photometry is particularly noticeable above [Fe/H] = $-$3.5 and this feature has been reported by \citet{Howes:2015aa}, \citet{Jacobson:2015aa} and \citet{Marino:2019aa}. Possible reasons were touched on above and explored in more detail in \citet{DaCosta:2019aa}. 

We now seek to compare the predicted and observed numbers of CEMP objects. Assuming uncorrected carbon abundances and the CEMP threshold of [C/Fe] $\ge$ +0.7, \citet{Placco:2014aa} report cumulative CEMP frequencies for [Fe/H] $\le$ $-$3.0, $-$3.5, and $-$4.0 of 32\%, 51\%, and 81\%, respectively. For our sample (including SMSS~J160540.18-144323.1), the numbers of stars in those three metallicity regimes are 76, 16, and 2. Therefore, we would expect 24.3, 8.2, and 1.6 CEMP objects in the metallicity ranges [Fe/H] $\le$ $-$3.0, $-$3.5, and $-$4.0, respectively, The numbers of CEMP stars in those metallicity ranges are 8, 4, and 2. Assuming Poisson statistics, for [Fe/H] $\le$ $-$3.0 the lack of CEMP stars compared to the predicted number is significant at the 2.9-$\sigma$ level. For [Fe/H] $\le$ $-$3.5, the difference between the predicted and observed numbers of CEMP stars represents only a 1.2-$\sigma$ result. Below [Fe/H] = $-$4.0, the statistics are small but the predicted and observed number of CEMP stars is in agreement. 

When adopting the corrected carbon abundances and corresponding predictions from \citet{Placco:2014aa}, the differences between the predicted and observed numbers of CEMP stars for [Fe/H] $\le$ $-$3.0 and $-$3.5 are significant at the 2.1-$\sigma$ and 0.6-$\sigma$ levels, respectively (including SMSS~J160540.18-144323.1). Therefore, any missing CEMP stars in our sample lie predominantly in the range [Fe/H] $>$ $-$3.5, which supports the discussion presented in \citet{DaCosta:2019aa}. We also note that \citet{Caffau:2020aa} reported a smaller fraction of CEMP stars in their sample with [Fe/H] $\le$ $-$2.5 (5\%; 3 out of 55) when compared to the \citet{Placco:2014aa} prediction of 19\% in that metallicity range. 

Nitrogen abundances were measured in some 20 objects in the current sample with metallicities in the range $-$4.1 $\le$ [Fe/H] $\le$ $-$2.3. The [N/Fe] ratios range from $-$0.5 to +2.6. Among the sample, we identify 11 objects which are nitrogen-enhanced metal-poor (NEMP) stars as defined by \citet{Johnson:2007aa} to have [N/Fe] $>$ +0.5 and [C/N] $<$ $-$0.5. One of these objects, SMSS~J030428.44-340604.8, was studied by \citealt{Hollek:2011aa} but they did not report a N abundance. Among these 11 NEMP stars, two are enriched in Eu (r-I) and one is a CEMP-no object (J054650.97-471407.9 with [Fe/H] = $-$4.09). 
The metallicity distribution of the 11 NEMP stars (median [Fe/H] = $-$3.27) does not appear to be in any way different from the overall metallicity distribution of the sample.  

Internal mixing could decrease C and increase N \citep{Spite:2005aa,Placco:2014aa}, which would reduce the NEMP fraction in our sample. If the corrected [C/N] ratio becomes $\sim 0$, then these stars could be explained by the same scenario for CEMP-no stars, i.e., faint supernovae; the C/N ratios decrease with stellar rotation or any hydrogen mixing during stellar evolution of the progenitor massive stars (e.g., \citealt{Nomoto:2013aa}). However, it is likely that our NEMP fraction remains high; for example, the evolutionary corrections to the observed C abundance for the NEMP stars J054650.97-471407.9 and J102410-082802.8 are negligible (see Table \ref{tab:c}) so that the NEMP status of these stars remains unaffected by evolutionary corrections. 

The NEMP population are believed to have formed via a similar process to the CEMP-$s$ stars, namely accretion from an asymptotic giant branch (AGB) companion. The difference, however, is that sufficiently massive AGB companions synthesize more nitrogen than carbon, although it remains unclear, due to limitations in the modelling (e.g., see review by \citealt{Karakas:2014ab}), whether extremely metal-poor massive AGB stars will also produce significant amounts of $s$-process material. We note, however, that none of our NEMP candidates show any evidence for potential $s$-process element enhancements. Specifically, the [Sr/Fe] and [Ba/Fe] abundance ratios for the NEMP stars are not distinguished from those for the bulk of the sample.

Based on modelling the mass distribution in AGB-star binaries, \citet{Johnson:2007aa} predicted between 12\% and 35\% of their sample to be NEMP (i.e., between two and seven stars in their sample of 21 objects), but instead found zero. Subsequent studies have confirmed that NEMP stars are rare, e.g., \citet{Simpson:2019aa} reported only 80 such objects among the many thousands of metal-poor stars in the SAGA database \citep{Suda:2008aa}. We are therefore somewhat surprised to find 11 NEMP stars among our sample; with only 20 objects with nitrogen measurements, the NEMP fraction is 55\% $\pm$ 21\%. It is unclear why we have been so successful in finding these rare stars. Given the low S/N near the NH lines at 3360\AA, we were unable to measure N for the majority of our sample. Therefore, the 11 NEMP stars is a lower limit. \citet{Johnson:2007aa} predict NEMP fractions between 12\% and 35\%, and our value is consistent with the upper range of the predictions. 

\subsection{Na to Ba} 

\begin{figure*}
	\includegraphics[width=.80\hsize]{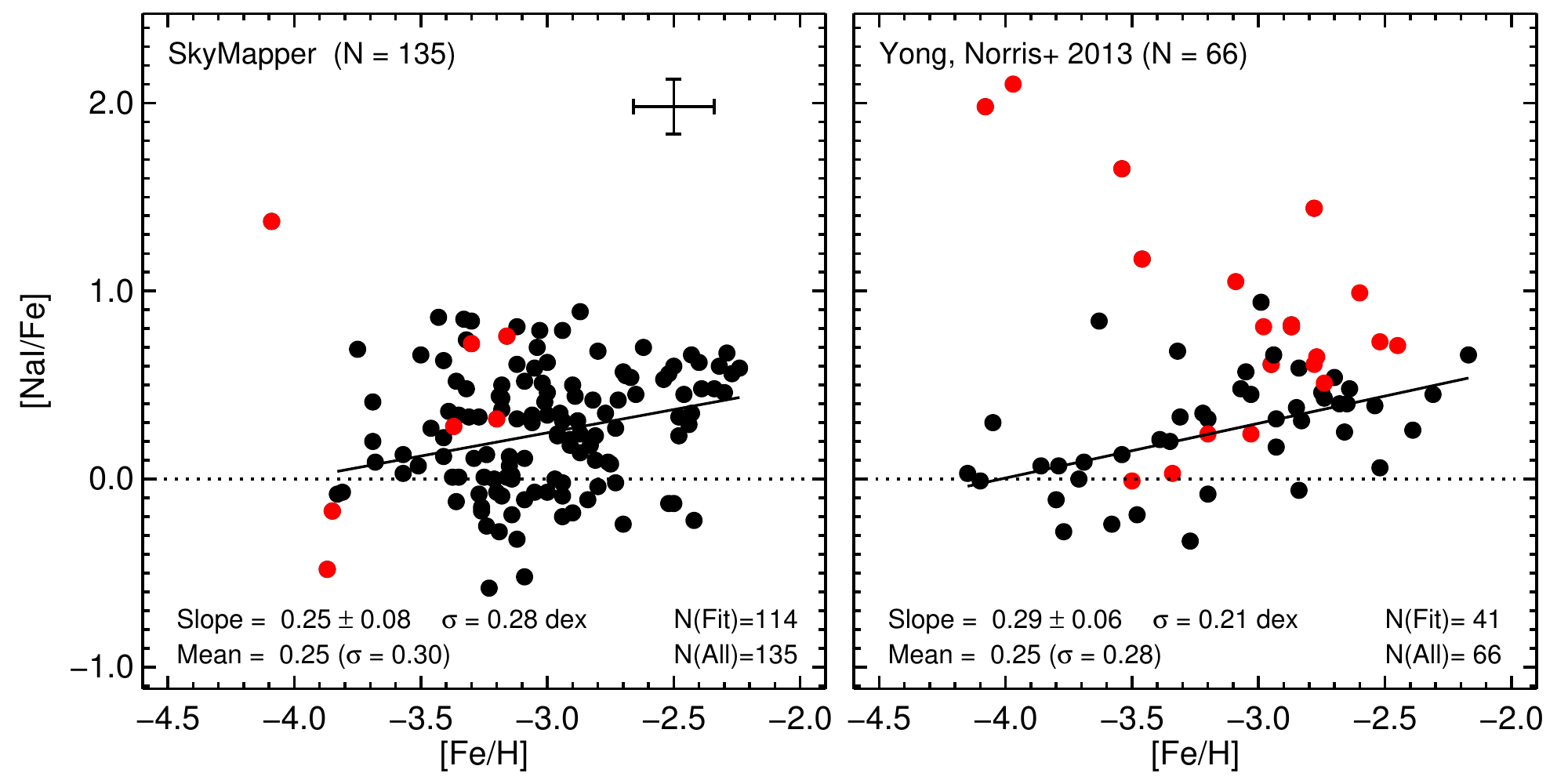}
    \caption{Same as Figure \ref{fig:C} but for Na. The linear fit to the data excluding C-rich stars and 2-$\sigma$ outliers is included in each panel along with the slope, uncertainty, dispersion about the linear fit, mean, and standard deviation.}
    \label{fig:Na}
\end{figure*}

\begin{figure*}
	\includegraphics[width=.80\hsize]{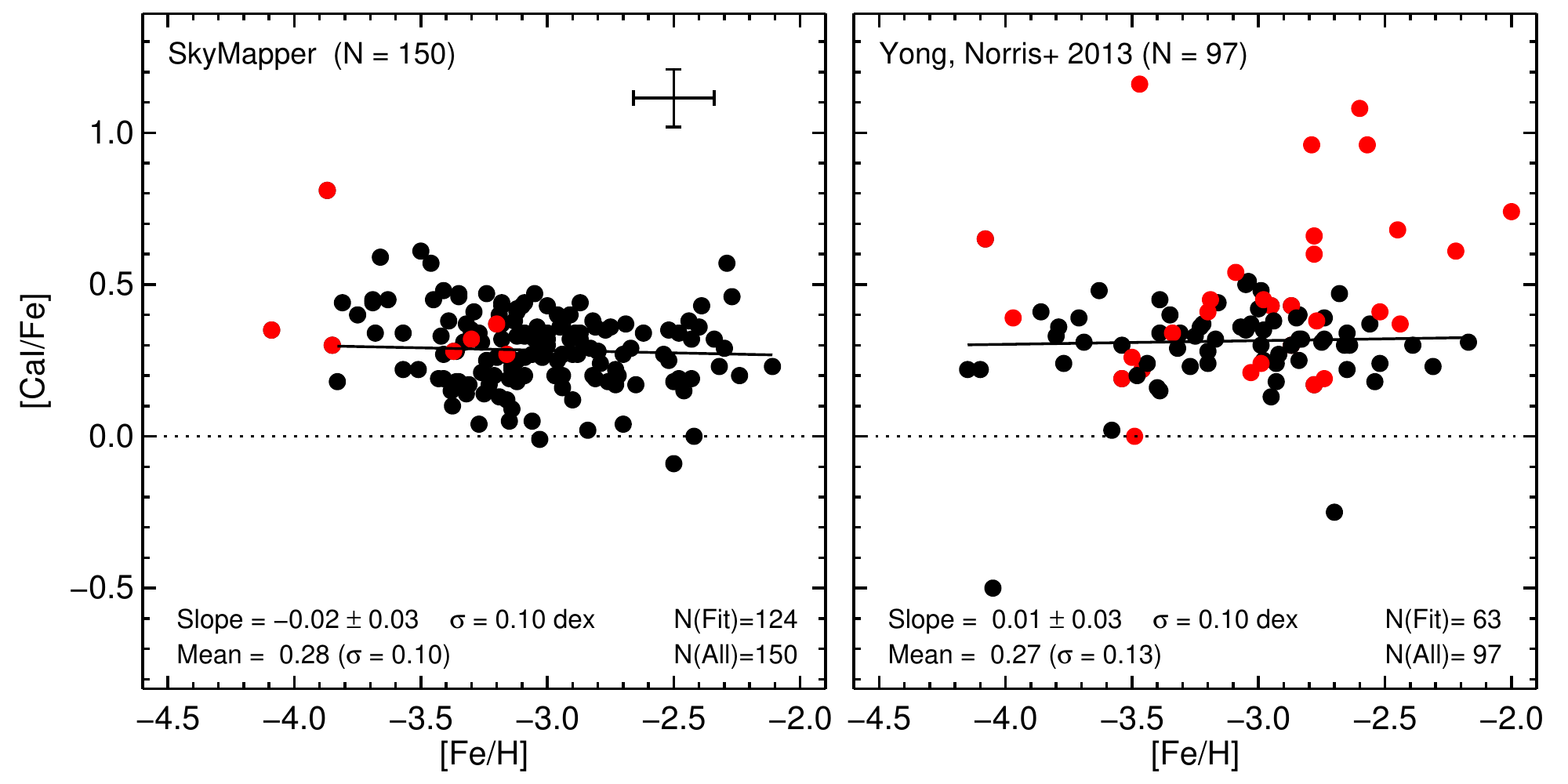}
    \caption{Same as Figure \ref{fig:Na} but for Ca.}
    \label{fig:Ca}
\end{figure*}

\begin{figure*}
	\includegraphics[width=.80\hsize]{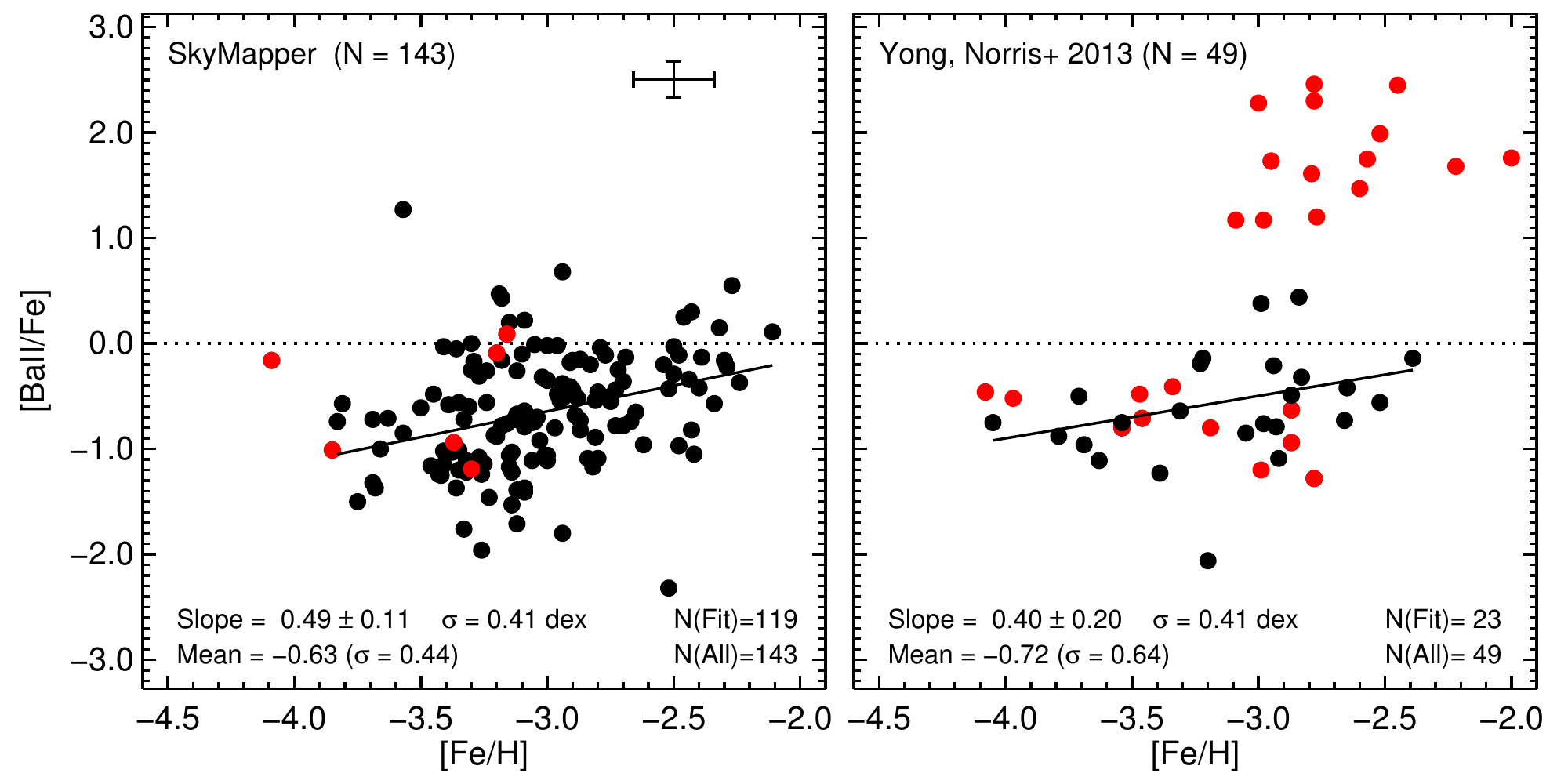}
    \caption{Same as Figure \ref{fig:Na} but for Ba.}
    \label{fig:Ba}
\end{figure*}

For the elements from Na to Ba, we plot the abundance distributions for three representative elements Na (odd-$Z$ element), Ca ($\alpha$ element) and Ba (neutron-capture element) in Figures \ref{fig:Na}, \ref{fig:Ca} and \ref{fig:Ba}. In each panel, we include the linear fit to the data excluding C-rich objects and 3-$\sigma$ outliers. The slope and uncertainty of the linear fit, the dispersion about the fit, the mean and standard deviation are included in each figure. In the right panels of those figures, we also include the giant stars from the \citet{Norris:2013ab} and \citet{Yong:2013ab} sample. For all elements, the current sample exhibits very similar behaviour to the \citet{Norris:2013ab} sample. 

\begin{figure*}
	\includegraphics[width=.99\hsize]{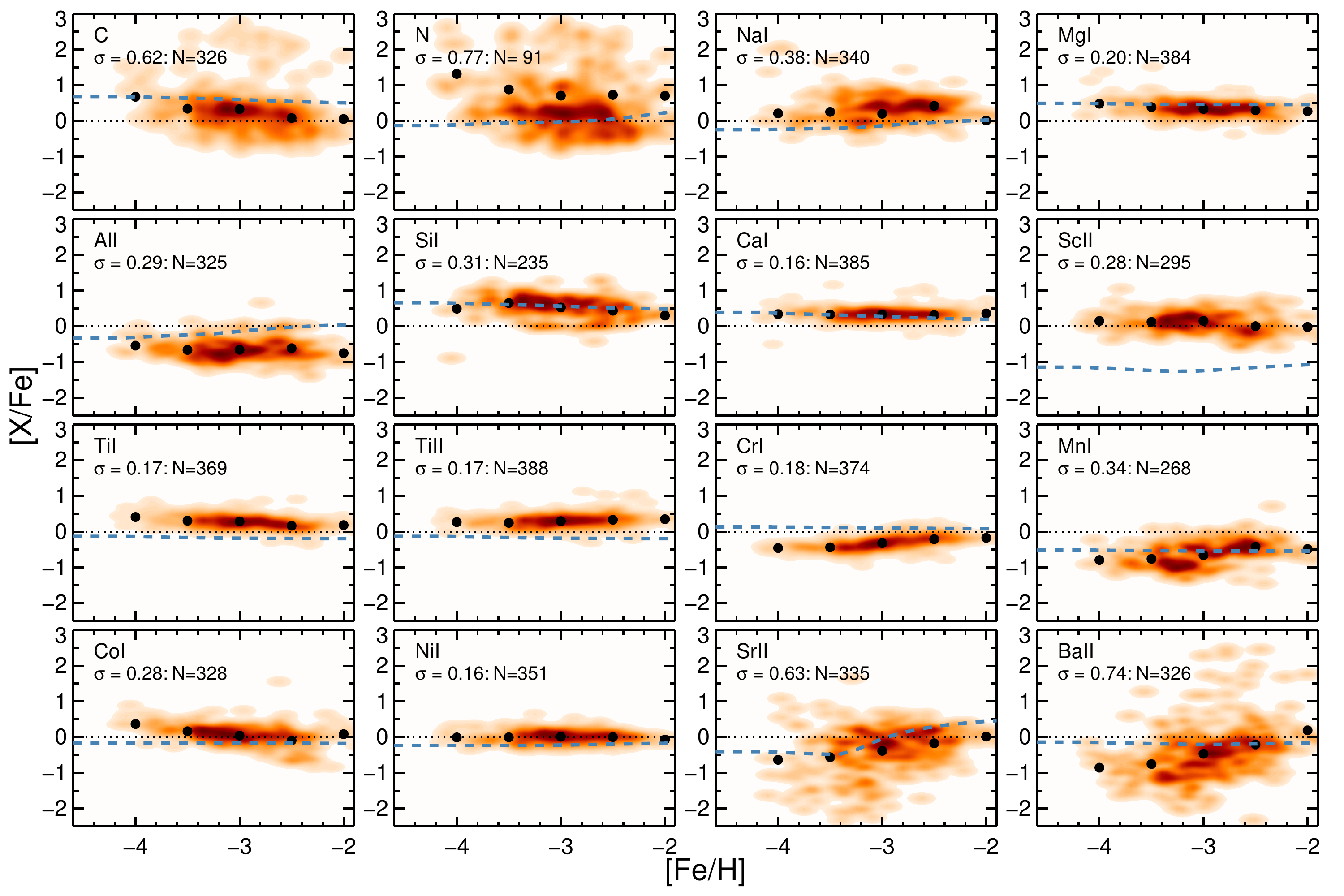}
    \caption{[X/Fe] vs. [Fe/H] for the elements from C to Ba for the current SkyMapper sample, \citet{Norris:2013ab}, \citet{Jacobson:2015aa} and \citet{Marino:2019aa}. (Only stars with [Fe/H] $\le$ $-$1.5 and \logg\ $<$ 3 are included.) We represent the data using generalized histograms of width 0.15 dex for [Fe/H] and 0.30 dex for [X/Fe]. Average values are overplotted as filled circles and the dashed blue lines are predictions from a Galactic chemical evolution model by \citet{Kobayashi:2020aa} updated to include stellar rotation (Kobayashi et al.\ in prep).}
    \label{fig:abunall}
\end{figure*}

In Figure \ref{fig:abunall}, we combine the current sample with \citet{Norris:2013ab}, \citet{Jacobson:2015aa} and \citet{Marino:2019aa} where we plot stars with [Fe/H] $\le$ $-$1.5 and \logg\ $<$ 3. We again represent the data using generalised histograms of width 0.15 dex for [Fe/H] and 0.30 dex for [X/Fe]. The total sample includes 479 stars of which 220 lie below [Fe/H] = $-$3, 128 have [Fe/H] $\le$ $-$3.25, 56 have [Fe/H] $\le$ $-$3.5, 29 have [Fe/H] $\le$ $-$3.75 and 10 have [Fe/H] $\le$ $-$4.0, although not every element is measured in every star. While there are selection biases associated with each of the individual samples, we believe that this figure provides the most extensive view to date of the early chemical enrichment of the Milky Way Galaxy. The filled circles plotted at [Fe/H] = $-$4, $-$3.5, $-$3.0, $-$2.5 and $-$2.0 in each of the panels of Figure \ref{fig:abunall} represent the average [X/Fe] ratios in the ranges, 
[Fe/H] $\le$ $-$3.75, 
$-$3.75 $<$ [Fe/H] $\le$ $-$3.25, 
$-$3.25 $<$ [Fe/H] $\le$ $-$2.75, 
$-$2.75 $<$ [Fe/H] $\le$ $-$2.25 and 
[Fe/H] $>$ $-$2.25, respectively. 
We present those values as columns A, B, C, D and E in Table \ref{tab:summary}. 

\begin{table}
	\centering
	\caption{Mean [X/Fe] values in various metallicity ranges for the combined sample shown in Figure \ref{fig:abunall}. 
	(A: 	[Fe/H] $\le$ $-$3.75. 
	B: $-$3.75 $<$ [Fe/H] $\le$ $-$3.25. 
	C: $-$3.25 $<$ [Fe/H] $\le$ $-$2.75. 
	D: $-$2.75 $<$ [Fe/H] $\le$ $-$2.25. 
	E: 	[Fe/H] $>$ $-$2.25.) [updated Mar 21]
	\label{tab:summary}}
	\begin{tabular}{lrrrrr}
	\hline
	Element & 
	 \multicolumn{5}{|c|}{Mean [X/Fe]} \\
	  &
	A &
	B &
	C &
	D & 
	E \\
	\hline
C        &       0.68 &       0.34 &       0.33 &       0.08 &       0.05  \\
N        &       1.32 &       0.88 &       0.71 &       0.73 &       0.70  \\
  \nai   &       0.21 &       0.25 &       0.20 &       0.42 &       0.01  \\
  \mgi   &       0.48 &       0.39 &       0.33 &       0.30 &       0.27  \\
  \ali   &    $-$0.54 &    $-$0.66 &    $-$0.66 &    $-$0.61 &    $-$0.75  \\
  \sii   &       0.49 &       0.65 &       0.53 &       0.44 &       0.30  \\
  \cai   &       0.34 &       0.33 &       0.33 &       0.31 &       0.36  \\
 \scii   &       0.15 &       0.12 &       0.15 &       0.00 &    $-$0.02  \\
  \tii   &       0.41 &       0.31 &       0.29 &       0.17 &       0.18  \\
 \tiii   &       0.27 &       0.25 &       0.29 &       0.33 &       0.35  \\
  \cri   &    $-$0.46 &    $-$0.44 &    $-$0.32 &    $-$0.21 &    $-$0.17  \\
  \mni   &    $-$0.79 &    $-$0.76 &    $-$0.66 &    $-$0.42 &    $-$0.49  \\
  \coi   &       0.36 &       0.16 &       0.04 &    $-$0.10 &       0.08  \\
  \nii   &    $-$0.01 &    $-$0.01 &       0.01 &    $-$0.00 &    $-$0.07  \\
 \srii   &    $-$0.64 &    $-$0.56 &    $-$0.39 &    $-$0.17 &       0.01  \\
 \baii   &    $-$0.86 &    $-$0.76 &    $-$0.47 &    $-$0.21 &       0.19  \\
	\hline
	\end{tabular}
\end{table}

It is clear from that figure that elements such as Ca, Ti, Cr and Ni have significantly smaller dispersion when compared to C, N, Na, Sr and Ba. For the former elements, the standard deviation for [X/Fe] in Figure \ref{fig:abunall} is 0.16 to 0.18 dex. For the latter elements, the standard deviation is $>$0.35 dex. In particular, for the neutron-capture elements Sr and Ba, the $\sim$3 dex range in [X/Fe] at low metallicities is well known (e.g., \citealt{McWilliam:1998aa},  \citealt{Roederer:2013aa}). We note that the average uncertainties range from 0.05 dex (Cr and Ni) to 0.15 dex (Sr), that is, the observed standard deviation exceeds the average measurement uncertainty. The standard deviations for the other elements depicted lie between those for Ca, Ti, Cr and Ni and those for C, N, Sr and Ba, with typical values of 0.28 dex. 

We note that while the trends in Figure \ref{fig:abunall} are similar to those presented in other studies, \citet{Cayrel:2004aa} and \citet{Reggiani:2017aa} have achieved higher abundance precision and smaller dispersion when using higher quality spectra. 

In Figure \ref{fig:abunall} we also include predictions from a Galactic chemical evolution model by \citet{Kobayashi:2020aa} (updated to include stellar rotation, Kobayashi et al.\ in prep) as the dashed blue lines (these represent the average values in their solar neighbourhood model). In general, there is reasonable agreement between the observational data (1D LTE analysis) and the theoretical predictions. Notable exceptions include N, Sc and Ti which are known to exhibit differences between theory and observations. The detailed discussion of the differences between the observations and the model can be found in \citet{Kobayashi:2020aa}, but briefly: the differences for Na, Al, Cr, and Co are potentially due to NLTE effects. Those for N and Ba can be solved with the inclusion of stellar rotation (e.g., \citealt{Kobayashi:2011ab}, Kobayashi et al.\ in prep). Those for Sc and Ti might be solved with multi-dimensional effects, although there are no successful explosion models that can be used in galactic chemical evolution models.

\subsection{Zn, Y, Zr and Eu} 

We now consider Zn, Y, Zr and Eu. In the previous subsections, we excluded these elements since we were unable to measure abundances in the majority of stars. (As noted earlier, Zn, Y, Zr and Eu abundances were only measured in 35, 23, 27 and 26 stars, respectively.) 

\begin{figure}
	\includegraphics[width=.90\hsize]{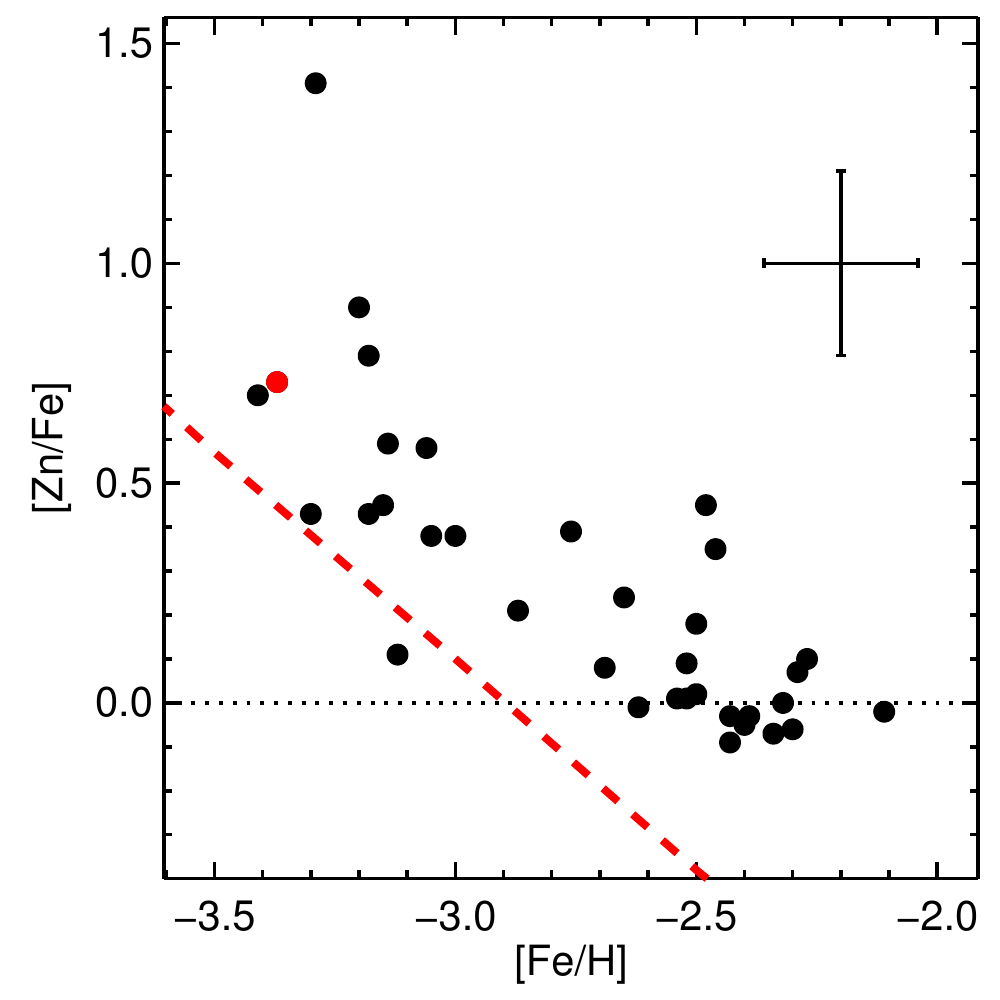}
    \caption{[Zn/Fe] vs.\ [Fe/H]. The red symbol is a C-rich star. The red dashed line indicates the approximate detection threshold for the 4810\AA\ \zni\ line.}
    \label{fig:zn}
\end{figure}

In Figure \ref{fig:zn}, there is a clear increase in [Zn/Fe] with decreasing [Fe/H]. That trend, however, is a direct consequence of the detection threshold for Zn. Following \citet{Roederer:2013aa}, we computed approximate abundance thresholds corresponding to the detection of the 4810\AA\ \zni\ line for a representative giant star with \teff\ = 4500 K and \logg\ = 1.5 for metallicities ranging from [Fe/H] = $-$5.0 to $-$2.0 (in steps of 0.5 dex). The red dashed line in the left panel of Figure \ref{fig:zn} corresponds to a line strength of 10m\AA. For example, at [Fe/H] = $-$3.5 a 10 m\AA\ \zni\ 4810.53\AA\ line would correspond to roughly [Zn/Fe] = +0.7. Basically for a fixed \teff\ and \logg, at lower metallicities Zn measurements are only possible as the [Zn/Fe] ratio increases. The 10m\AA\ value and stellar parameters were arbitrarily chosen and we regard those abundances as approximate detection thresholds. Nevertheless, the abundance dispersion appears to increase with decreasing metallicity. One object, SMSS~J084327.83-141513.3, exhibits an unusually high Zn abundance with [Zn/Fe] = +1.4. We will briefly discuss this object in the following subsection. 

\color{black}

\begin{figure*}
    \includegraphics[width=.80\hsize]{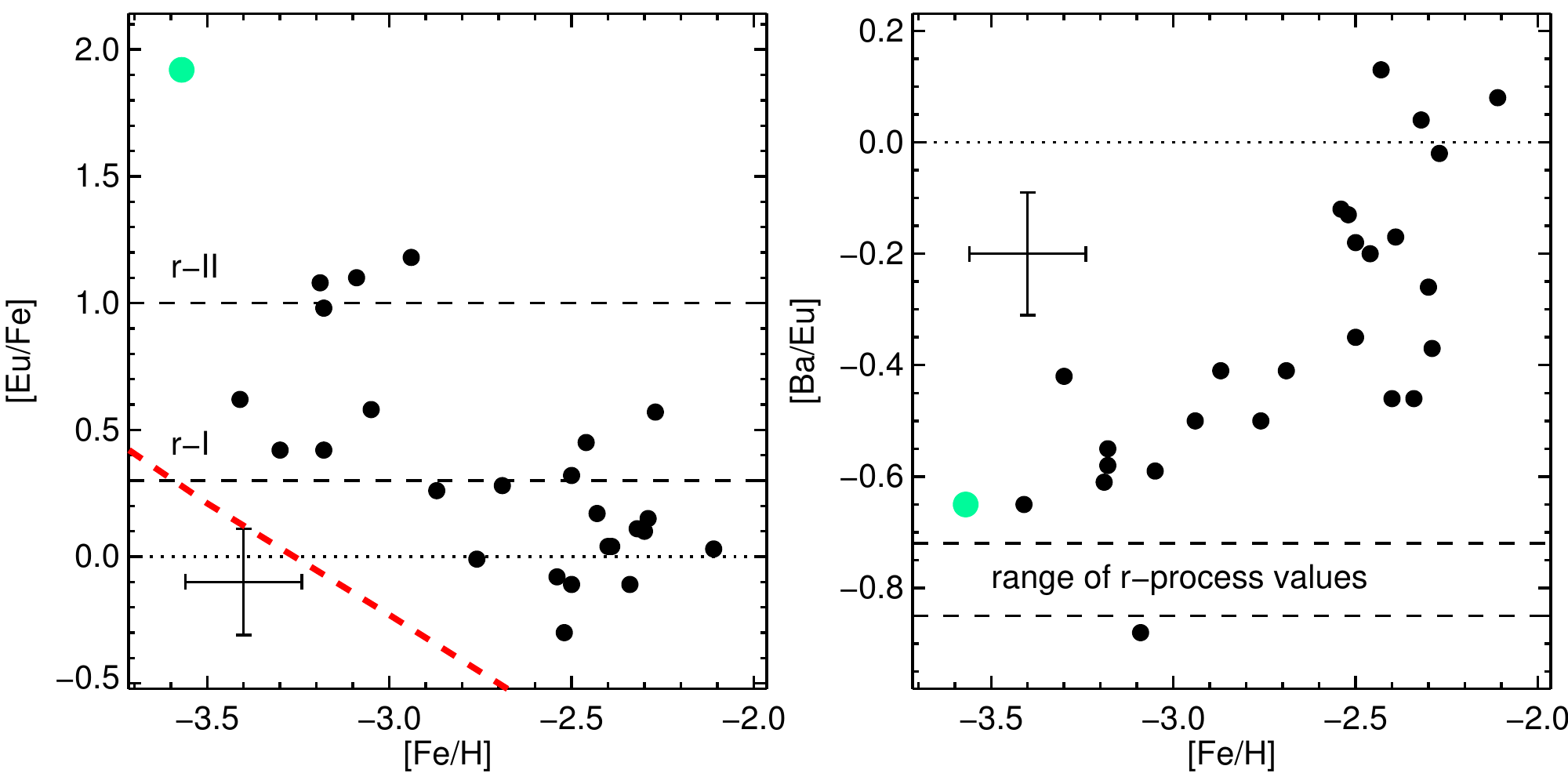}
    \caption{(Left) [Eu/Fe] vs.\ [Fe/H]. The aqua circle represents the neutron-capture rich star SMSS~J200322.54-114203.3 and a representative error bar is included in the bottom left corner. The red dashed line indicates the approximate detection threshold for the 4129\AA\ \euii\ line. (Right) [Ba/Eu] vs.\ [Fe/H]. The range scaled solar r-process values ([Ba/Eu] = $-$0.72 to $-$0.85) are indicated as dashed lines.}
    \label{fig:Eu}
\end{figure*}

In Figure \ref{fig:Eu} (left panel), we plot [Eu/Fe] as a function of metallicity. While the data indicate a trend of increasing [Eu/Fe] with decreasing [Fe/H], we are mindful that this is again due in part to the detection threshold for Eu. As described above for Zn, we followed the approach of  \citet{Roederer:2013aa} and computed approximate abundance thresholds corresponding to the detection of the 4129\AA\ \euii\ line.  

We identify four new r-II stars ([Eu/Fe] $>$ +1.0 and [Ba/Eu] $<$ 0) corresponding to 2.7\% of our sample which is consistent with previous studies of the frequency of such objects, $\sim$3\%  \citep{Barklem:2005aa,2018ARNPS..68..237F}. We also report eight new r-I stars (0.3 $\le$ [Eu/Fe] $\le$ 1.0 and [Ba/Eu] $<$ 0) which represents 5.3\% of our sample which is lower than the $\sim$15\% previously noted \citep{Barklem:2005aa,2018ARNPS..68..237F}. In calculating these fractions we have assumed that the stars for which the 4129\AA\ \euii\ line is not detected are not Eu-rich. Given the estimated detection threshold for this line shown in Figure \ref{fig:Eu}, this does not seem an unreasonable assumption, particularly since the r-I and r-II fractions for only stars in which Eu is detected (8/26 r-I; 4/26 r-II) would seem unreasonably high compared to the observed fractions in other surveys \citep[e.g.,][]{Barklem:2005aa,2018ARNPS..68..237F}

Among these Eu rich objects, we note that the r-I stars SMSS~J110901.23+075441.7 and  SMSS~J181200.10-463148.8 are also NEMP objects. All other r-I and r-II stars are neither CEMP nor NEMP\footnote{A more extensive chemical abundance analysis of the most iron-poor r-II star SMSS~J200322.54-114203.3 using a new spectrum is presented in \citet{Yong:2021aa}. In that analysis, SMSS~J200322.54-114203.3 is found to also be a NEMP object. Co-existing $r$-process and N enhancements might indicate an origin involving massive rotating stars and their supernovae.}. 

In Figure \ref{fig:Eu} (right panel), we plot [Ba/Eu] against [Fe/H]. The data show an increase in [Ba/Eu] with increasing metallicity and this is possibly due to the contribution from the slow neutron-capture process \citep{McWilliam:1998aa}. The lowest values of [Ba/Eu] are consistent with the solar system $r$-process values which range from $-$0.85 to $-$0.72 \citep{Kappeler:1989aa}. In both panels, we highlight the location of SMSS~J200322.54-114203.3 which is highly enhanced in neutron-capture process elements. A comprehensive abundance analysis of this object is presented in \citet{Yong:2021aa}. 

\subsection{Chemically peculiar stars}  

\subsubsection{Outliers}

\begin{figure*}
	\includegraphics[width=.85\hsize]{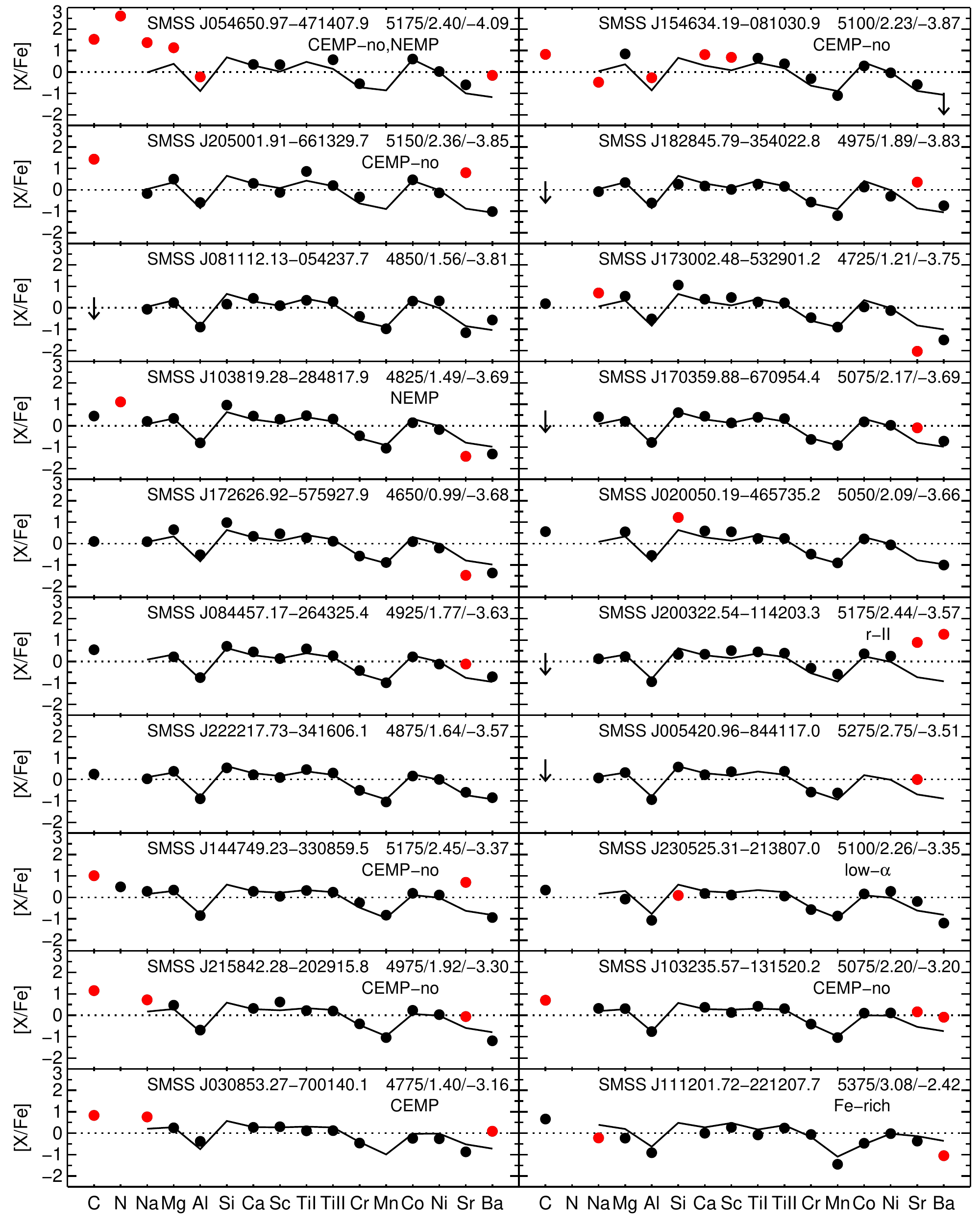}
    \caption{Abundance patterns [X/Fe] for each element for 20 program stars that are CEMP and/or with [Fe/H] $<$ $-$3.5 along with  representative ``Fe-rich'' and ``low-$\alpha$'' objects ordered by increasing metallicity. In each panel, the solid line represents the ``normal'' [X/Fe] ratio at the metallicity of each program star. The model parameters (\teff, \logg, [Fe/H]) are included in each panel. Red circles indicate when [C,N/Fe] $\ge$ +0.7 dex, or when [X/Fe] differs from the solid line by more than 0.5 dex.}
    \label{fig:individual}
\end{figure*}

In Figure \ref{fig:individual}, we plot the abundance pattern [X/Fe] for each element for 20 stars in our sample that are CEMP and/or with [Fe/H] $<$ $-$3.5 as well as one ``Fe-rich'' star and one low-$\alpha$ star. The solid line in each panel indicates the [X/Fe] ratio that a ``normal'' star would have at the metallicity of each program star. That abundance is defined using the linear fit to [X/Fe] versus [Fe/H] for which three examples are illustrated in Figures \ref{fig:Na} through \ref{fig:Ba}. 

The red points in Figure \ref{fig:individual} indicate elements which depart (above or below) from the solid line by more than 0.5 dex. For C and N, the red points indicate [X/Fe] $\ge$ +0.7 dex (no evolutionary corrections were applied). We regard the red points as being peculiar with respect to the [X/Fe] ratio that a normal star would have at the metallicity of the program star. This system enables us to readily identify which elements may be regarded as outliers for a given star, and therefore which stars are chemically peculiar, i.e., they exhibit multiple elements that are outliers. 
As noted earlier, six of the seven CEMP stars belong to the CEMP-no subclass, i.e., they have [Ba/Fe] $<$ 0. The remaining CEMP star SMSS~J030853.27-700140.1, with [Ba/Fe] = +0.09 $\pm$ 0.17, cannot be placed into any sub-class but when taking into account the error bars, this object could also be a CEMP-no object. As discussed in \citet{Norris:2013aa}, CEMP objects often exhibit high abundance ratios for Na, Mg and Al. Among the seven CEMP objects in Figure \ref{fig:individual}, we note that four exhibit unusually high abundances for Na, Mg and/or Al. However, for one of those objects, SMSS~J154634.19-081030.9, the high Al abundances are accompanied by a low Na abundance. While the other three CEMP stars have normal Na, Mg and Al abundances, they exhibit high Sr and/or Ba abundances when compared to the average star at the same metallicity, although  
they are not CEMP-$s$ which have [Ba/Fe] $>$ +1.0. 
\color{black}

There are a further 13 stars which are CEMP when taking into account the evolutionary correction for carbon \citep{Placco:2014aa}. All belong to the CEMP-no subclass except for SMSS~J054903.50-594655.4 which has [Ba/Fe] = +0.30 dex.  This star does not fall into any standard CEMP subclass as the Ba abundance lies between the CEMP-no and CEMP-$s$ thresholds of [Ba/Fe] $<$ 0 and $>$ +1, respectively. Among these objects, we note that two are also NEMP stars (SMSS~J103819.28-284817.9 and SMSS~J190836.24-401623.5) and one (SMSS~J173002.48-532901.2) has a high Na abundance. 

As noted in \citet{Yong:2013aa}, peculiar element abundance ratios are also commonly found among the C-normal population. Among the C-normal stars (i.e., excluding CEMP with and without the evolutionary correction) at all metallicities, some 75 have at least one element from Na to Ba for which the [X/Fe] ratio differs from a normal star at the same metallicity by at least 0.5 dex (above or below). If we exclude Sr and Ba, which are known to exhibit an enormous range in [X/Fe] at low metallicity, 
there are some six C-normal stars which are chemically peculiar, but only SMSS~J232121.57-160505.4 has more than one element that is peculiar (in this case Na and Al). Given the large sample at low metallicity, these stars that exhibit peculiar chemical abundances may represent examples of stochastic chemical enrichment. 

\subsubsection{Low-$\alpha$}

We identify a subset of stars which exhibit low [$\alpha$/Fe] ratios where $\alpha$ is the average of Mg, Si, Ca, \tii\ and \tiii. The nine objects with [$\alpha$/Fe] $\le$ +0.15 are listed in Table \ref{tab:alpha}, and we note that all are C-normal. For comparison, the \citet{Norris:2013ab} and \citet{Yong:2013ab} sample includes 11 giants (out of 98) with [$\alpha$/Fe] $\le$ +0.15 and [Fe/H] $\le$ $-$2.0. The frequency of low-$\alpha$ stars in this study, 6.0 $\pm$ 2.1\%, is smaller than in \citet{Yong:2013ab}, 11.2 $\pm$ 3.6\% although the difference is only at the 1$\sigma$ level. 

\begin{table}
	\centering
	\caption{Stars with [$\alpha$/Fe] $\le$ +0.15 where $\alpha$ is the average of Mg, Si, Ca, \tii\ and \tiii\ and $\sigma$ is the standard deviation. (As discussed in the text, the first seven stars are ``Fe-rich'' while the final two are ``$\alpha$-poor''.) [updated Nov 20]}
	\label{tab:alpha}
	\begin{tabular}{lcrc}
	\hline
	ID &
	[Fe/H] & 
	$<$[$\alpha$/Fe]$>$ & 
	$\sigma$[$\alpha$/Fe] \\ 
	&
	(dex) &
	(dex) &
	(dex) \\
	\hline
SMSS~J011126.27-495048.4  &  $-$2.94 &     0.06 &   0.07  \\
SMSS~J031703.94-374047.2  &  $-$3.27 &     0.13 &   0.15  \\
SMSS~J034749.80-751351.7  &  $-$2.50 &     0.04 &   0.32  \\
SMSS~J093524.93-715506.5  &  $-$2.48 &     0.13 &   0.16  \\
SMSS~J111201.72-221207.7  &  $-$2.42 &  $-$0.02 &   0.20  \\
SMSS~J170133.47-651115.6  &  $-$2.70 &     0.03 &   0.03  \\
SMSS~J190508.31-581843.9  &  $-$3.19 &     0.11 &   0.08  \\
SMSS~J230525.31-213807.0  &  $-$3.35 &     0.06 &   0.11  \\
SMSS~J232121.57-160505.4  &  $-$3.03 &     0.12 &   0.25  \\
	\hline
	\end{tabular}
\end{table}

Adopting the same approach as in Figure \ref{fig:individual}, we can explore whether these nine objects appear deficient in just the $\alpha$ elements or whether the [X/Fe] ratios are generally low when compared to other stars at similar metallicity. In particular, we can compare the [X/Fe] ratio in each star with that of the ``typical'' star at the same metallicity. For seven of these nine objects, the deficiency in $\alpha$ elements is essentially identical to the deficiency in all elements from Na to Ni. We therefore propose that these are ``Fe-enhanced'' stars as discussed in \citet{Cayrel:2004aa}, \citet{Venn:2012aa}, \citet{Yong:2013aa} and \citet{Jacobson:2015aa}. That is, the unusually low [X/Fe] ratios in these objects can most simply be explained as an excess of Fe. For these seven objects, removing $\sim$0.20 dex of Fe would result in [X/Fe] ratios that are typical for stars at the same metallicity. 

In the study of \citet{Cordoni:2020aa}\footnote{
Relative to the values presented in \citet{Cordoni:2020aa}, minor updates to the abundances have been made, but none of the conclusions of that paper are affected.} the kinematics of these seven  ``Fe-enhanced'' stars are quite diverse, likely indicating a variety of origins. Two stars are members of the inner Galactic halo, one is an outer-halo star, one is likely associated with the Gaia-Enceladus-Sausage accretion event \citep{Helmi:2018aa,Belokurov:2018aa}, two have prograde orbits in the thick-disk and one has high energy and is escaping from the Galaxy.

For the other two objects, SMSS J230525.31-213807.0 and SMSS J232121.57-160505.4, the deficiency in $\alpha$ elements is considerably greater than for all elements from Na to Ni. For SMSS J230525.31-213807.0, the [$\alpha$/Fe] ratio differs by $-$0.29 dex from that for the average star at the same metallicity while for all elements from Na to Ni, the [X/Fe] differ by $-$0.12 dex. For SMSS J232121.57-160505.4, the same two quantities are $-$0.16 and +0.10 dex. We therefore propose that these two stars are $\alpha$-poor and not ``Fe-enhanced''. Both of these stars have prograde orbits that are largely confined to the Galactic Disk in the kinematic analysis of  \citet{Cordoni:2020aa}: SMSS~J230525.31-213807.0 has an orbital eccentricity of 0.58, peri- and apo-galactic distances of 2.2 kpc and 8.4 kpc respectively, and a maximum excursion from the Galactic plane of 2.5 kpc, while for SMSS~J232121.57-160505.4, the corresponding numbers are 0.28, 5.9 and 10.4 kpc and 1.4 kpc, respectively. Given the low [Fe/H] and low [$\alpha$/Fe] values for these stars (see Table \ref{tab:alpha}), we speculate that they most likely originated in now disrupted low-luminosity (ultra-faint) dwarf galaxies. (See \citealt{Kobayashi:2014aa} for a more detailed discussion of the origins of low-$\alpha$ metal-poor stars.) 

\subsubsection{SMSS~J200322.54-114203.3}

As noted above SMSS~J200322.54-114203.3 is the most iron-poor r-II star known: [Fe/H] = $-$3.57 (based on \fei\ lines) and [Eu/Fe] = +1.92. A more comprehensive chemical abundance analysis of this object is presented in \citet{Yong:2021aa} in which the $r$-process enrichment is attributed to magnetorotational hypernovae. 

\subsubsection{SMSS~J084327.83-141513.3}

As noted above, SMSS~J084327.83-141513.3 with [Fe/H] = $-$3.29 is particularly enhanced in Zn with [Zn/Fe] = +1.41. For the neutron-capture elements, it has [Sr/Fe] = +1.64, [Y/Fe] = +0.69, [Zr/Fe] = +1.02 and [Ba/Fe] = $-$0.17. (The Eu limit is [Eu/Fe] $<$ +0.57.) Inspection of the SAGA database \citep{Suda:2008aa} indicates that among the stars with [Fe/H] $<$ $-$3 it has the highest [Zn/Fe] and [Sr/Fe] ratios by about 0.20 dex, and we note that HE 1327-2326, with [Fe/H]$_{\rm NLTE}$ = $-$5.2, also exhibits large Zn and Sr enhancements \citep{2019ApJ...876...97E}. The [Sr/Ba] ratio is +1.8 which is even more extreme than the value of +1.0 reported for the $\omega$ Centauri star ROA 276 \citep{Yong:2017aa} in which the abundance pattern could be explained by spinstars, i.e., fast-rotating low-metallicity massive stars \citep{Frischknecht:2012aa,Frischknecht:2016aa}. For SMSS~J084327.83-141513.3, however, we consider the possibility that the enrichment arises from an electron-capture supernova (ECSN); a more extensive analysis of this object will be presented elsewhere (Nordlander et al.\ in prep). 

\subsubsection{SMSS~J165501.84-664110.7}

This object is relatively metal-rich with [Fe/H] = $-$2.52 with a very low value of [Ba/Fe] = $-$2.32. In Figure \ref{fig:Ba} it notable for being about 1 dex lower in [Ba/Fe] when compared to stars at the same metallicity. However, it has only a moderately low value of [Sr/Fe] = $-$0.82 and appears otherwise normal in [X/Fe] ratios for all other elements except for Al where the observed and ``normal star'' values are [Al/Fe] = $-$0.08 and $-$0.65, respectively. 

\section{Conclusions}

We present chemical abundances for 21 elements from Li to Eu for a sample of 150 stars selected from the SkyMapper survey spanning $-$4.1 $<$ [Fe/H] $<$ $-$2.1. Our study is based on high-resolution, high S/N spectra adopting a 1D LTE analysis. Our sample includes 90 stars with [Fe/H] $\le$ $-$3, seven CEMP (and a further 13 when including the evolutionary correction), 11 NEMP (at least two remain NEMP if we include evolutionary mixing corrections), eight r-I and four r-II objects. Of those seven CEMP stars, six belong to the CEMP-no subclass and the other cannot be assigned to any particular subclass. One of the CEMP-no objects is also a NEMP star, and all CEMP objects lie below [Fe/H] = $-$3.0. Two of the NEMP stars are also r-I and one is also a CEMP-no object (SMSS~J054650.97-471407.9). 

We combine our sample with previous studies \citep{Norris:2013ab,Yong:2013ab,Nordlander:2019aa} for which there are some 177 stars with [Fe/H] $\le$ $-$3. The metallicity distribution function has a slope of $\Delta$(log N)/$\Delta$[Fe/H] = 1.51 dex per dex in the range $-$4 $\le$ [Fe/H] $\le$ $-$3 which is comparable to the value of 1.5 $\pm$ 0.1 dex per dex in \citet{DaCosta:2019aa} but steeper than the value of 1.0 from the  \citet{Hartwick:1976aa} simple model. If we exclude CEMP objects, the MDF has a slope of 1.74 $\pm$ 0.02. When considering the metallicity range $-$3.4 $<$ [Fe/H] $<$ $-$2.7, we find the slope of the MDF is +1.07 $\pm$ 0.04 which is similar to the value of +1.0 $\pm$ 0.1 found by \citet{Youakim:2020aa} over the range $-$3.4 $<$ [Fe/H] $<$ $-$2.5 in the Pristine survey. Both the present study and that of \citet{Youakim:2020aa} find a marked turn down in the MDF at [Fe/H] $\approx$ --3.8; more metal-poor objects are predominantly carbon-rich.

We find that the chemical abundance ratios [X/Fe] as a function of [Fe/H] exhibit similar trends to those noted in the literature. There are two stars that are particularly unusual, and more comprehensive chemical abundance analyses are presented elsewhere: SMSS~J200322.54-114203.3 is highly enhanced in the $r$-process elements and the abundance pattern could be explained by magnetorotational hypernovae \citep{Yong:2021aa}; SMSS J084327.83-141513.3 has the highest [Zn/Fe] and [Sr/Fe] ratios among all stars with [Fe/H] $<$ $-$3 in the SAGA database and could be explained by enrichment from an electron-capture supernova (Nordlander et al.\ in prep). 

Overall, our large and homogeneously analysed sample of metal-poor stars is a substantial contribution towards a better understanding of chemical enrichment at the earliest times. The key to constraints on the properties of the first generation of zero-metallicity stars, however, lies with the extremely rare objects for which [Fe/H] $<$ $-$4; the search for such stars remains on-going. 

\section*{Acknowledgements}

Australian access to the Magellan Telescopes was supported through the National Collaborative Research Infrastructure Strategy of the Australian Federal Government. The authors wish to recognize and acknowledge the very significant cultural role and reverence that the summit of Maunakea has always had within the indigenous Hawaiian community. We are most fortunate to have the opportunity to conduct observations from this mountain. We thank V.\ Placco for providing evolutionary corrections for carbon and A.\ I.\ Karakas for helpful comments on AGB nucleosynthesis. We thank the referee for helpful comments. 

Parts of this research were supported by the Australian Research Council Centre of Excellence for All Sky Astrophysics in 3 Dimensions (ASTRO 3D), through project number CE170100013. 
GDC acknowledges Australian Research Council grant DP150103294. 
KL acknowledges funds from the European Research Council (ERC) under the European Union's Horizon 2020 research and innovation programme (Grant agreement No. 852977). 
ADM acknowledges Australian Research Council grant FT160100206. 
AFM acknowledges support from the European Union's Horizon 2020 research and innovation programme under the Marie Sklodowska-Curie grant agreement No 797100. 
ARC acknowledges Australian Research Council grant DE190100656. 
CK acknowledges funding from the UK Science and Technology Facility Council (STFC) through grant ST/R000905/1 \& ST/V000632/1, and the Stromlo Distinguished Visitor Program at the ANU.

\section*{Data Availability} 

The data underlying this article will be shared on reasonable request to the corresponding author.




\bibliographystyle{mnras}






\bsp	
\label{lastpage}
\end{document}